\documentclass{aa}  

\usepackage{graphicx}
\usepackage{txfonts}
\usepackage{hyperref}
\usepackage{color}
\usepackage{soul}
\usepackage{orcidlink}

\hypersetup{hidelinks}

\begin{document}

    \defcitealias{collaboration_desi_2016-1}{DESI Collaboration}
    \defcitealias{collaboration_desi_2016}{DESI Collaboration}
    \defcitealias{collaboration_euclid_2024}{Euclid Collaboration}
    \defcitealias{collaboration_euclid_2024-1}{Euclid Collaboration}
    \defcitealias{blanchard_euclid_2020}{Euclid Collaboration}
    \defcitealias{collaboration_euclid_2025}{Euclid Collaboration}
    \defcitealias{aghanim_planck_2020}{Planck Collaboration}

   \title{Dynamic zoom simulations of structure formation beyond standard cosmology}

   \titlerunning{Dynamic zoom simulations of structure formation beyond standard cosmology}

   \author{Riccardo Zangarelli\inst{1,2}\fnmsep\thanks{\email{riccardo.zangarelli2@unibo.it}}\orcidlink{0009-0001-0302-9680}, Marco Baldi \inst{1,2,3}\orcidlink{0000-0003-4145-1943}, Federico Marinacci\inst{1,2}\orcidlink{0000-0003-3816-7028}, Enrico Garaldi\inst{4,5}\orcidlink{0000-0002-6021-7020}}

   \authorrunning{R. Zangarelli et al.}

   \institute{Dipartimento di Fisica e Astronomia ``Augusto Righi'', Universit\`a di Bologna, via Piero Gobetti 93/2, I-40129 Bologna, Italy.
   \and
   INAF, Osservatorio di Astrofisica e Scienza dello Spazio di Bologna, via Piero Gobetti 93/3, I-40129 Bologna, Italy.
   \and
   INFN, Sezione di Bologna, viale Berti Pichat 6/2, I-40127 Bologna, Italy.
   \and
   Kavli IPMU (WPI), UTIAS, The University of Tokyo, Kashiwa, Chiba 277-8583, Japan.
   \and
   Center for Data-Driven Discovery, Kavli IPMU (WPI), UTIAS, The University of Tokyo, Kashiwa, Chiba 277-8583, Japan.
   }

   \date{\today}
 
  \abstract
   {A thorough interpretation of the current and upcoming generation of cosmological observations requires unprecedented large-scale, high-resolution simulations that span multiple cosmological models and parameters. These simulations are extremely computationally demanding, and their realization poses a crucial technical challenge.}
   {We present beyond-$\Lambda$CDM implementations of the dynamic zoom simulations (DZSs) method, a performance-enhancing technique tailored for large-scale simulations that produce light-cone-like outputs. This approach dynamically decreases the resolution of a simulation in those regions that are not in causal connection with the observer, which substantially saves computational resources without directly affecting the physical properties within the light cone.}
   {We implemented the DZS algorithm in two state-of-the-art codes that support nonstandard cosmologies, namely modified $f(R)$ gravity in Arepo and dark sector interactions in Gadget4. We then analyzed the result accuracy and performance gains across resolution, simulation volume, and cosmological model by comparing simulations performed with and without the DZS algorithm.}
   {We show that our DZSs reproduce the light cone halo mass function, sky-projected massmaps, and matter and weak lensing convergence power spectra with an accuracy of $\simeq 0.1\%$ or higher in most cases. In terms of performance, DZSs in our tests can save up to $\sim 50\%$ runtime compared to the non-DZS counterparts. Moreover, scaling our results to larger simulated volumes suggests that performance gains could improve by an additional $\sim 20\%$ at the resolution levels of current state-of-the-art cosmological simulations.}
   {The validation of the DZS algorithm in beyond-$\Lambda$CDM models demonstrates that this technique can enable cost-effective, large-scale ($\gtrsim 1$ cGpc$/h$) simulations with state-of-the-art resolution, and provide the computational framework needed to constrain and help the interpretation of forthcoming observational data.}

   \keywords{Cosmology --  Methods: numerical -- Cosmology: theory -- (Cosmology:) dark matter -- (Cosmology:) dark energy -- (Cosmology:) large-scale structure of Universe
}

   \maketitle
   \nolinenumbers

\section{Introduction} \label{sec:intro}

Cosmological simulations have become a fundamental tool in modern astrophysics. They make it possible to conduct ``virtual experiments'' on large portions of the Universe, by exploring different cosmologies across various models and parameter spaces. Indeed, the physical processes that shape galaxies and the large-scale structure (LSS) are highly nonlinear, which severely limits the applicability of purely analytical approaches and instead requires the use of numerical tools. This means that simulations can be effectively viewed as a theoretical counterpart to real observations \citep[e.g.,][]{springel_trends_2012, somerville_models_2015, angulo_reviews_2022, primack_galaxy_2024}.

Recently, observations have become more accurate and deeper than ever thanks to state-of-the-art instruments, such as \textit{Euclid} (\citealt{laureijs_euclid_2011}; \citetalias{collaboration_euclid_2025} \citeyear{collaboration_euclid_2025}), the \textit{Vera C. Rubin} Observatory \citep[][]{ivezic_lsst_2019}, and the Dark Energy Spectroscopic Survey (\citetalias{collaboration_desi_2016-1} \citeyear{collaboration_desi_2016-1, collaboration_desi_2016}), and upcoming ones such as the Square Kilometer Array \citep{braun_advancing_2015}, Athena \citep{nandra_hot_2013}, and the \textit{Nancy Grace Roman }Space Telescope \citep{spergel_wide-field_2015}. For a thorough interpretation of this massive amount of data, simulations need to keep up with the unprecedented quality and sky coverage of current observations, providing high-resolution views of wide volumes of the Universe and modeling a large statistical sample of objects with increasing physical fidelity. The latter requirement would ideally imply that the key physical processes that drive the formation of observable (luminous) objects -- such as gas cooling, star formation, feedback mechanisms (both from stars and active galactic nuclei), and magnetic fields, to name a few -- are self-consistently included in the simulation. However, full-physics simulations with a mass resolution $\lesssim 10^{9}$ M$_\odot/h$ on gigaparsec scales are extremely expensive (when feasible at all) with the current generation of high-performance computing facilities, especially when different models or parameters need to be explored through multiple realizations. 

A faster and computationally less intensive approach is represented by $N$-body simulations, coupled with post-processing techniques for galaxy modeling -- such as the halo occupation distribution model \citep[][]{berlind_halo_2002}, subhalo abundance matching \citep[][]{conroy_modeling_2006}, and semi-analytical models (\citealt{henriques_galaxy_2015, somerville_star_2015, hirschmann_galaxy_2016, lacey_unified_2016}).
Nevertheless, not even the $N$-body scenario is free of technical challenges. Hundreds of billions or even trillions of simulation particles (i.e., discrete tracers of the underlying continuous fields) are needed to cover a large, gigaparsec-scale volume with a somewhat high mass resolution ($\lesssim10^{10}$ M$_\odot/h$), resulting in huge memory, storage space, and computational requirements. Consequently, it should not come as a surprise that some of the largest simulations to date (e.g.,  \textit{Euclid} Flagships I and II, \citealt{potter_pkdgrav3_2017}; \citetalias{collaboration_euclid_2024} \citeyear{collaboration_euclid_2024};  Millennium-XXL, \citealt{angulo_scaling_2012};  DEUS Full Universe, \citealt{alimi_deus_2012}) all required millions of core-hours on $\sim 10^4$ computational units.

Moreover, the typical simulation output format (a so-called snapshot of the whole volume at a fixed cosmic time) does not reflect the survey view of the Universe, as in the latter we receive information from different cosmic times due to the finite value of the speed of light. While a combination of snapshots can mimic this light-cone-like representation -- for instance through a piecewise-constant approximation between snapshots adjacent in time -- this approach still requires us to save to disk and post-process a substantial volume of data, as snapshots can weigh several terabytes each in the most expensive simulations. Therefore, simulations have recently started to switch to an on-the-fly light-cone-output approach, either by saving particles in thin shells with a radius equal to the light-travel distance, $R_{lc}$, \citep{evrard_galaxy_2002, fosalba_onion_2008, potter_pkdgrav3_2017}, or with more sophisticated light-cone-crossing interpolations of individual particles \citep{springel_simulating_2021}. These on-the-fly techniques eliminate the need for a large amount of ``traditional'' snapshots for large simulations, with the main output now being the light-cone-like one (a few snapshots might still be saved at representative redshifts for further analysis, e.g.,\ on 3D clustering). However, the light-cone-output approach does not mitigate the intensive use of computational resources required in modern large-scale cosmological simulations.

This light-cone-like view raises the concern of having an increasingly large fraction of the simulated volume that sits outside of $R_{lc}(t)$: as the latter decreases with increasing time, $t$, more and more particles cross the light cone and are no longer in causal connection with the observer. Moreover, at low redshift, the nonlinear clustering of matter requires substantially more computational resources than at higher redshifts. The combination of nonlinearity and a decreasing $R_{lc}$ implies that eventually most of the computational time and memory available to a simulation will be employed to evolve structures that are discarded from the main output. Therefore, finding a way to reduce the resources used outside of the light cone represents a logical approach to mitigate the high computational toll taken by state-of-the-art simulations.

A practical implementation of this idea is given by the shrinking domain framework \citep[SDF;][]{llinares_shrinking_2017} where the gravitational calculations are only carried out for particles that lie within the light cone; after a particle crosses it, it is discarded from the simulation. The SDF is very effective in reducing the computational time taken by a simulation, as numerical calculations involve fewer particles with decreasing redshift. However, many gravity solving algorithms in cosmological simulations \citep[including the one employed in][]{llinares_shrinking_2017} rely on the periodicity of the simulated volume. Therefore, substantial modifications to the numerical methods are needed for the SDF to produce correct results. Most importantly, it should also be noted that cosmological simulations are typically carried out within the Newtonian framework of gravity, where changes in the gravitational potential propagate instantaneously.\footnote{Normally, this approximation yields accurate modeling of even nonlinear perturbations, because their scale is well inside the cosmological horizon (equivalently, their velocity is far below the speed of light). Therefore, matter clustering takes place in a non-relativistic regime, which is safe to model with Newtonian dynamics \citep{chisari_newtonian_2011}.} Consequently, gravitational interactions on scales larger than the light cone are not modeled self-consistently in  simulations that adopt the SDF technique.

Due to these shortcomings, it would be preferable to still perform gravitational calculations outside of the light cone, but at a reduced resolution with respect to the region within $R_{lc}$. This reduced resolution can be achieved by progressively merging simulation particles outside of the light cone, to yield more massive discrete tracers that are still part of the simulation (unlike in the SDF). In this way, fewer computational resources are spent in a simulation, without any major modifications to the $N$-body solver or significant effects on the main light-cone-like output. This approach, dubbed dynamic zoom simulations (DZSs), was originally introduced in the code Gadget3 \citep[based on its predecessor Gadget2 described in][]{springel_cosmological_2005} by \cite{garaldi_dynamic_2020}. Their implementation managed to reduce the time taken by a simulation by up to $50\%$, while retaining $\sim 0.1\%$ accuracy, in $\Lambda$ cold dark matter (CDM), dark-matter-only setups. These results indicate that the DZS technique may offer a promising solution to mitigate the technical challenges of the next generation of cosmological simulations. While the Gadget3 implementation was limited to the $\Lambda$CDM framework -- and therefore did not allow for the modeling of alternative cosmologies -- our work enables the use of various models across more modern codes.

In this paper, we present upgraded implementations of the DZS technique in the state-of-the-art codes Arepo \citep[][]{springel_e_2010, weinberger_arepo_2020} and Gadget4 \citep[][]{springel_simulating_2021} for cosmological models beyond  $\Lambda$CDM, in order to allow a more performance-friendly analysis of different cosmologies besides the standard one. Specifically, we focus on $f(R)$ modified gravity \citep[MG, implemented in Arepo by][]{arnold_realistic_2019} and dark scattering \citep[DS; originally implemented in Gadget3 by][now available in Gadget4 as well]{baldi_simulating_2015}. The paper is structured as follows. A description of and physical motivation for these models and their implementations, along with details of the DZS technique, is presented in Sect. \ref{sec:models}. In Sect. \ref{sec:validation} we assess the performance gains of our implementations in dark-matter-only scenarios across the different cosmological models, as well as their accuracy in reproducing cosmologically relevant output data. Section \ref{sec:HR} evaluates the performance of the DZS algorithm in simulations with a higher resolution than those used in our limited testing setups. Finally, we present our conclusions in Sect. \ref{sec:conclusions}.

\section{Numerical methods and cosmological models} \label{sec:models}

\begin{figure*}[t]
    \includegraphics[width=\linewidth]{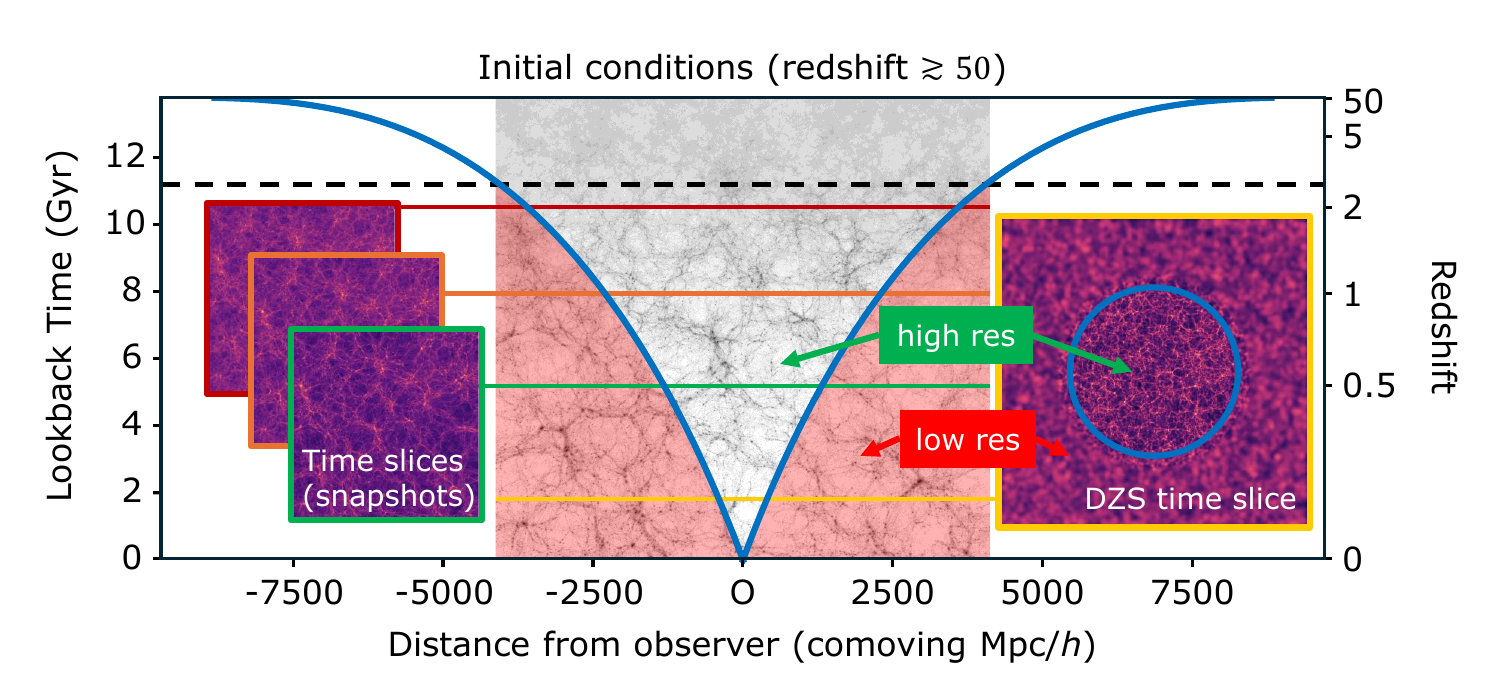}
    \caption{1D+1D space-time diagram illustrating the light-cone-like approach of DZSs (adapted from Fig.~1 of \citealt{llinares_shrinking_2017} and \citealt{garaldi_dynamic_2020}). An example simulation (a cubic box $8192$ comoving Mpc$/h$ on a side) is depicted from its initial conditions (top) to redshift, or lookback time, zero (bottom). The depicted density field is only included for display purposes, and does not match the size of the volume marked on the $x$ axis \citep[it is rendered from TNG300-3-Dark simulation data, ][]{nelson_illustristng_2019}. The observer is placed at redshift zero at the location marked with $O$. The simulation volume is crossed by the light cone (blue curve) at a time indicated by the dashed black line. Traditional snapshots, or time slices, are taken at the cosmic times highlighted by horizontal colored lines, which leads to fixed-time density maps of real simulation volumes, rendered from  the TNG300-3-Dark (left) and the \textsc{medium} DZS on the right (see Table~\ref{tab:simulations} in Sect. \ref{sec:validation}).}
    \label{fig:scheme}
\end{figure*}

The core idea behind performance enhancement methods such as DZSs and the SDF is that when the main output of a large-scale simulation is in the form of a light cone, the use of computational resources should not be evenly distributed across the simulation volume, but focused mostly on regions that are causally connected to the observer. This concept is depicted graphically in Fig. \ref{fig:scheme}, which is a 1D+1D space-time diagram of an example cosmological simulation, with the spatial dimension on the $x$ axis and the temporal one on the $y$ axis. The observer is placed at point $O$ at redshift zero, and the past light cone (blue curve) converges toward them from the initial conditions as the simulation progresses. During this process, the light cone will enter the simulated volume at some redshift $z_{lc}$, marked by the black dashed line. From $z_{lc}$ onward, an increasingly large fraction of the volume is left out of the light cone (red shaded area). While traditional snapshots, or ``time slices,'' include that volume (i.e., the red, orange, and green line in Fig.~\ref{fig:scheme} cross the red-shaded area), a light-cone-like output would only consist of the overlap between the blue light cone curve and the simulation volume. Therefore, the red shaded area can be discarded (as in the SDF) or adjusted on-the-fly to a lower resolution (as in DZSs). Time slices can still be extracted to illustrate this procedure, as shown by the yellow line in Fig.~\ref{fig:scheme}: the light cone -- marked by a blue circle with radius $R_{lc}$ in the time slice -- separates the inner high-resolution area from the outer low-resolution one. This resolution reduction operation in DZSs requires only minimal modifications to simulation codes, and no modification to their $N$-body solver. This aspect is a major advantage of the original DZS implementation over the SDF \citep{garaldi_dynamic_2020}. For the purpose of this work, we note that these properties also enable a relatively straightforward implementation of DZSs in the majority of modern cosmological simulation codes employing both $\Lambda$CDM and alternative models. For this reason, we implemented and present the DZS algorithm in codes that incorporate nonstandard gravity or exotic dark energy models, as modeling such scenarios is essential for accurately interpreting survey data.

As mentioned in the introduction, a class of nonstandard gravity models, the so-called $f(R)$ gravity, was already included in Arepo by \cite{arnold_realistic_2019}. Therefore, we chose to use this code to test and showcase the capabilities of the DZS algorithm with a nonstandard gravitational solver. Moreover, recent results from \cite{lodha_extended_2025} predict significant deviations from the $\Lambda$CDM picture through a redshift-dependent equation of state parameter for dark energy, $w_{DE}(z)$. For this reason, we also implemented DZSs in another code capable of modeling beyond-$\Lambda$CDM effects, namely PANDA-Gadget4 (Baldi \& Casalino, in prep.). Among other nonstandard scenarios, the code allows us to follow DS effects, which take place if $w_{DE} \neq w_{DE,\Lambda CDM} = -1$.
 
In what follows, we describe the main features of the two codes adopted in this work and the modeling, scientific motivation, and numerical implementations of nonstandard, beyond $\Lambda$CDM effects. We conclude this methodological section by detailing the key features of DZS implementations in Arepo and Gadget4.

\subsection{Arepo} \label{sec:arepo}
Arepo is a $N$-body and galaxy formation code, first presented in \cite{springel_e_2010} and publicly released in \cite{weinberger_arepo_2020}. It is massively parallel and employs the message passing interface \citep[MPI;][]{mpif_mpi_1993}, making it suited for state-of-the-art simulations distributed over a large number of computational units. Examples of these include the IllustrisTNG suite \citep{nelson_illustristng_2019} and the MillenniumTNG project \citep[][]{pakmor_millenniumtng_2023}.

Arepo follows the evolution of discrete mass tracers that interact gravitationally and, if gas is included, hydro-dynamically. Hybrid numerical schemes are used for both types of interactions, with a \texttt{treePM} algorithm handling gravity and a moving Voronoi mesh for hydrodynamics. Our current implementation of DZSs in Arepo supports $N$-body, collisionless simulations, so we focus only on the numerical treatment of gravitational interactions. We refer the interested reader to \cite{springel_e_2010} and \cite{weinberger_arepo_2020} for further details on the moving mesh approach.

The \texttt{treePM} scheme \citep[][]{xu_new_1995, bode_tree_2000, bagla_treepm_2002} combines the particle-mesh (PM)\ method \citep[e.g.,][]{klypin_three-dimensional_1983}, to compute long-range interactions, with the hierarchical multipole method \citep[or \texttt{``tree''} method;][]{barnes_hierarchical_1986} for short-range forces. In the following we briefly describe the implementation of the gravity tree, since it is crucial for both the DZS algorithm and the $f(R)$ gravity implementation described below.

The tree method approximates the contributions of distant groups of particles (the so-called ``nodes'' of the tree) through their multipole expansion, deciding which nodes are distant enough through a ``geometrical'' criterion -- although other, not distance based, criteria exist. The tree construction algorithm splits the cubic simulation domain, called ``root node,'' into eight equal-volume sub-domains, which are the ``children'' nodes of the root node. Each child node is again split into eight sub-domains, and the process continues recursively until the smaller nodes contain at most one simulation particle. This ensures that the tree naturally adapts to the gravity-driven clustering of particles, creating increasingly smaller nodes. Each node built during this construction process stores the mass and center of mass of its particle content (used to calculate multipole moments), as well as links to other nodes (i.e., its ``father,'' ``siblings,'' and ``children''). We anticipate that this tree structure can be employed in DZSs to reduce the simulation resolution outside of the light cone, because the properties of a node, i.e., quantities evaluated over its particle content, provide a straightforward way to merge those particles into more massive, and therefore lower-resolution, tracers. Furthermore, the presence of multiple, nested grids in the structure of the tree makes it well suited to be directly employed in multi-grid solvers, such as the one used by \cite{arnold_realistic_2019} to calculate modified gravity contributions to particle interactions.

\subsection{$f(R)$ gravity in Arepo} \label{sec:mgArepo}
The use of general relativity (GR) as a description of gravity is a fundamental assumption of the standard cosmological model. Several small-scale experimental tests agree with the predictions of GR up to high levels of precision \citep[][]{will_confrontation_2014}. Nonetheless, on scales larger than our Solar System GR remains largely unconstrained. 
For this reason, generalizations of GR that only influence cosmological scales (employing ``screening'' mechanisms to ensure that GR predictions are restored in high-density environments, where these have been tightly constrained) represent an interesting approach to achieve a better understanding of the nature of gravity and its role in the cosmic accelerated expansion and the evolution of the LSS. This type of investigations is particularly timely in view of forthcoming large-scale tests of gravity, enabled by precise measurements of object clustering from missions such as \textit{Euclid} (\citealt{laureijs_euclid_2011}; \citetalias{collaboration_euclid_2025} \citeyear{collaboration_euclid_2025}).

Among these generalizations, we focus on $f(R)$ gravity, where the Einstein field equations of GR are modified through a function ($f$) of the Ricci scalar ($R$). Specifically, we consider a class of $f(R)$ models that can account for the accelerated expansion of the Universe without needing a cosmological constant $\Lambda$ \citep[][]{hu_models_2007}. In the Newtonian approximation commonly employed in cosmological simulations, including $f(R)$ gravity effects yields a modified Poisson equation with the form
\begin{equation}
    \nabla^2 \Phi = \frac{16 \pi G}{3} \delta \rho - \frac{1}{6} \delta R \text{ ,}
    \label{eq:mg_poisson}
\end{equation}
where $\Phi$ is the gravitational potential, $G$ is the gravitational constant, and $\delta \rho$ and $\delta R$ are perturbations in the matter density and scalar curvature, respectively. The latter can be conveniently expressed in terms of $f_R \equiv \mathrm{d}f(R)/\mathrm{d}R$, which represents the scalar field responsible for beyond-GR effects. The approximated field equation for $f_R$ is \citep[e.g.,][]{oyaizu_nonlinear_2008}
\begin{equation}
    \nabla^2 f_R = \frac{1}{3} (\delta R - 8 \pi G \delta \rho) \text{ .}
    \label{eq:fR}
\end{equation}
Combining Eqs.~(\ref{eq:mg_poisson}) and (\ref{eq:fR}), the modification to standard gravity can be expressed in terms of an additional acceleration $\Vec{a}_{MG}$ experienced by a particle:
\begin{equation}
    \Vec{a}_{MG} = \frac{c^2}{2} \nabla f_R \text{ ,}
\end{equation}
which can be computed by solving Eq.~(\ref{eq:fR}). This additional contribution to gravitational interaction enhances them up to a factor of $4/3$. It can be shown \citep[][]{puchwein_modified-gravity-gadget_2013, arnold_realistic_2019} that the $f(R)$ gravity models by \cite{hu_models_2007} can be effectively parametrized by $\Bar{f}_{R0}$, i.e., the mean value of $f_R$ at redshift zero. Since these models are derived from a general broken power-law parametrization, its exponent, $n$, is also needed to describe them. Nonetheless, most studies, including \cite{arnold_realistic_2019}, set $n = 1$.

A computational treatment of $f(R)$ gravity has the main task of solving numerically Eq.~(\ref{eq:fR}). The implementation of \cite{arnold_realistic_2019} in Arepo (dubbed MG-Arepo) closely follows the MG-Gadget project of \cite{puchwein_modified-gravity-gadget_2013}: in both cases, the oct-tree from the \texttt{tree} N-body solver is used as an adaptive grid to solve a discretized version of Eq.~(\ref{eq:fR})\footnote{The main difference between the two codes lies in a different parametrization to express a discretized version of Eq.~(\ref{eq:fR}), although MG-Arepo reaches higher accuracy.}. This discretized version, however, is highly nonlinear, requiring the use of multi-grid iterative solvers. Nonetheless, even with the use of a multi-grid technique, the $f(R)$ solver leads to a substantial increase in computing time in Arepo simulations with respect to a $\Lambda$CDM setup: this increase amounts to a factor of $2$ or more for setups analyzed in this work, but it can very well exceed one order of magnitude for higher-resolution runs (e.g., Appendix A in \citetalias{collaboration_euclid_2024-1} \citeyear{collaboration_euclid_2024-1}). This makes MG-Arepo the ideal example of a computationally heavy $N$-body framework that can benefit from performance gains brought by the DZS algorithm (see Sect. \ref{sec:validation} for details).

As a final remark, we would like to note that the version of MG-Arepo presented in \cite{arnold_realistic_2019} does not include light cone output capabilities, which are a key element of DZSs. We added this crucial feature by porting into MG-Arepo the Gadget4 implementation of the light cone output mode described in \citet[][]{springel_simulating_2021}. As already mentioned in the Introduction, this approach to the light cone output is a significant advancement with respect to the thin shell approximation employed in \cite{fosalba_onion_2008} and 
in the original version of the DZS algorithm as well \citep{garaldi_dynamic_2020}.

\subsection{Gadget4} \label{sec:gadget4}
Gadget4 \citep[][]{springel_simulating_2021} is a $N$-body and galaxy formation code which shares many similarities with Arepo in the handling of gravitational interactions, as it also employs the \texttt{treePM} scheme (generally with the same workflow described in Sect. \ref{sec:arepo}). Hydrodynamics, on the other hand, is modeled through the smoothed particle hydrodynamics (SPH) approach. We refer to \cite{springel_simulating_2021} for details on the SPH technique.

As well as Arepo and many other massively parallel codes, Gadget4 also adopts MPI as the parallelization framework, albeit in a hybrid shared-distributed memory variant. We see in Appendix \ref{sec:wl} that this detail is needed to understand the performance of our Gadget4 simulations.

\subsection{Dark scattering in Gadget4} \label{sec:DSGadget}
Findings from \cite{lodha_extended_2025} suggest a significant deviation of the dark energy equation of state from the time-independent behavior prescribed by $\Lambda$CDM. Coupled with our little understanding of the physical nature of dark matter and dark energy (the so-called dark sector), this leaves ample room for speculations about the physical framework that drives the evolution of these components. While a similar line of reasoning with respect to constraints on GR leads to the formulation of modified gravity theories (see Sect. \ref{sec:mgArepo}), a hypothetical interaction between the dark sector components might very well be nongravitational \citep[e.g.,][]{wetterich_asymptotically_1995, amendola_coupled_2000, farrar_interacting_2004}. In fact, the exchange of momentum and energy between dark matter and dark energy can also alleviate cosmological tensions, for example between early and late measurements of the clustering of matter \citep[e.g.,][]{pourtsidou_reconciling_2016, baldi_structure_2017}.

\begin{figure}[t]
   \resizebox{\hsize}{!}{\includegraphics{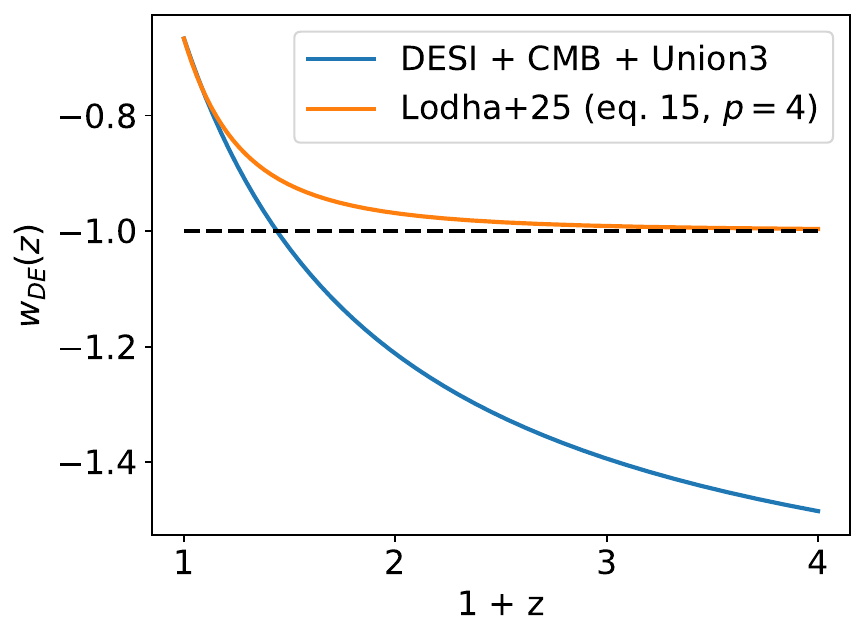}}
      \caption{Equation of state (EoS) parameter of dark energy, $w_{DE}$, as a function of redshift ($z$). The blue curve is the best-fit CPL parametrization from \cite{collaboration_desi_2025}, with $w_0 = -0.667$, $w_a = -1.09$. Its ``phantom'' behavior ($w_{DE} < -1$) at high $z$ is highlighted by the dashed black line, which marks $w_{DE} = -1$. The orange curve represents a thawing EoS (characterized by $w_{DE} \rightarrow -1$ for $z \rightarrow + \infty$) taken from \citet[][their  Eq.~15]{lodha_extended_2025}, and reproducing the low-$z$ trend of the CPL best-fit without the phantom behavior.} 
         \label{fig:wDE}
   \end{figure}

In this work, we focus on pure momentum exchange between dark matter and dark energy, i.e., an interaction akin to an elastic scattering. When modeling the dark sector components as a fluid, the exchange arises as an additional ``drag'' term $A(z)$ in the momentum equation, with the form \citep[e.g.,][]{simpson_scattering_2010, baldi_structure_2017, cruickshank_forecasts_2025}
\begin{equation}
    A(z) \equiv (1 + w_{DE}) \frac{\sigma}{m_{CDM}}\frac{3\Omega_{DE}(z)}{8 \pi G} H(z) \text{ ,}
    \label{eq:DSdrag}
\end{equation}
where $\sigma$ is the scattering cross section, $m_{CDM}$ is the mass of a cold dark matter particle (throughout this paper we use $\sigma/m_{CDM} = 100$ cm$^2/$GeV), $\Omega_{DE}$ is the dark energy density parameter, and $H(z)$ is the Hubble parameter. It is now clear that $A \neq 0$ only if $w_{DE} \neq -1$, that is, a nonstandard evolution of dark energy. We chose to adopt DESI-inferred trends for $w_{DE}(z)$, using the best-fit (DESI+CMB+Union3) values from \cite{collaboration_desi_2025}, namely a Chevallier-Polanski-Linder (CPL) parametrization $w_{DE}(z) = w_0 + w_a z/(1+z)$ with $w_0 = -0.667$, $w_a = -1.09$. This $w_{DE}(z)$ is plotted in Fig.~\ref{fig:wDE} (blue curve). Notably, this parametrization exhibits ``phantom crossing,'' that is, $w_{DE} < - 1$ at high $z$. Both the transition into this ``phantom regime'' and the regime itself pose serious theoretical challenges \citep[e.g.,][]{carroll_can_2003, vikman_can_2005}; therefore, we also employ a so-called ``thawing'' parametrization, where $w_{DE}$ is ``locked'' at $-1$ at high $z$ and is then allowed to move in the non-phantom regime ($w_{DE} \geq -1$) at lower redshifts.
We adopt the parameterization in \citet[][their  Eq.~15]{lodha_extended_2025} and choose $p=4$ to mimic the low-$z$ behavior of the best DESI+CMB+Union3 fit (as seen in the orange curve of Fig. \ref{fig:wDE}). In accordance to the thawing behavior, in this case we choose $w_a = -0.333$ \citep[$w_a$ appears in][their Eq.~15]{lodha_extended_2025}, so that $w_0 + w_a = -1$.

The numerical implementation of DS in Gadget4 is relatively straightforward. We employ the PANDA-Gadget4 framework (Baldi \& Casalino, in prep.), which reads tabulated values for $w_{DE}(z)$, $\Omega_{DE}(z)$, $\Omega_M(z)$ and $H(z)$ and uses them to calculate $A(z)$, which then yields the DS contribution to the acceleration of tracer dark matter particles. Unlike in MG-Arepo, the computational toll of this additional interaction is essentially negligible. However, in our DZS showcase framework, PANDA-Gadget4 proves that even when the nonstandard calculations are computationally light, DZSs can still yield significant performance improvements (as shown in Sect. \ref{sec:performance}).

\subsection{Dynamic zoom simulations algorithm}
The practical implementation of DZSs in both MG-Arepo and PANDA-Gadget4 builds on top of the oct-tree structure, as in the original Gadget3 implementation. The tree nodes contain the integrated physical quantities of their particle content, such as mass and center of mass. In fact, the gravitational approximation employed by the tree treats nodes as if they were a single ``fictitious'' particle, holding the combined mass of the particles inside the node and located in its center of mass. In the DZS algorithm, the nodes fulfilling a ``derefinement'' criterion are converted into actual particles and their particle content eliminated. In other words, when a tree node can be de-refined, DZS replaces its particle content with the corresponding single ``fictitious" particle, which is then turned into a real mass tracer.

More specifically, the DZS algorithm operates from a redshift $z_{lc}$, at which the light cone enters the volume of a simulation, by performing the following tasks in any global discrete timestep:\footnote{Both Arepo and Gadget4 employ discrete timesteps to advance the trajectories of simulated particles. Each particle is allowed to have its own timestep (e.g., for an accurate integration of high-density environments). The DZS technique can only operate when all particles adjust to the same simulation time, i.e., ``synchronize''. This happens in so-called ``global'' timesteps.}
\begin{itemize}
    \item calculate the value of $R_{lc}$ at the current step;
    \item perform a depth-first ``tree walk'' \citep[akin to the one done by the \texttt{tree} solver and detailed in][]{garaldi_dynamic_2020} to check if nodes outside of the light cone can be merged into a particle, according to a user-defined de-refinement criterion. Those nodes are flagged for de-refinement and the particles they contain are flagged for elimination;
    \item remove from the simulation the particles flagged for elimination;
    \item use the mass, center of mass, and center of mass velocity of flagged nodes to create new low-resolution particles to replace the ones that have been removed.
\end{itemize}
Repeating these operations over and over as the simulation advances yields a progressively reduced number of particles, as most of the simulation volume adjusts to a lower resolution. This results in decreased memory and run time requirements, without significant modifications to the light-cone-like output.

Besides the tree itself, the de-refinement criterion employed in DZSs can also be borrowed from the hierarchical multipole solver. In fact, the DZS adaptation of the geometric criterion tests the distance between a node located outside of the light cone and the "outer edge" of the light cone itself, i.e., the spherical surface with radius $R_{lc}$. If the node is far enough from the light cone edge, it satisfies the criterion and is flagged for de-refinement. More specifically, when
\begin{equation}
    \frac{L}{|\Vec{s}| - (R_{lc} + b)} < \theta_{geom}
    \label{eq:geom}
\end{equation}
a node is flagged and its particles are merged. Here, $\theta_{geom}$ is a user-defined threshold, $L$ is the linear  size of a node, $\Vec{s}$ is the position of its center of mass with respect to the observer, and $b$ is a user-defined ``buffer length'' which keeps an additional volume beyond the light cone at the original simulation resolution. This length ensures that at $z=0$ (where $R_{lc}=0$) a sphere with radius $b$ is kept at the original simulation resolution. Note that Eq.~(\ref{eq:geom}) only applies to nodes outside of the light cone, for which $|\Vec{s}| > R_{lc} + b$. More generally, a user-tweakable de-refinement criterion enables a gradual transition between the original and lower resolution, minimizing the impact of DZSs on the dynamics of particles inside the light cone and giving the user a high degree of control over DZS performance and accuracy. In all the simulations analyzed in this work, we set $\theta_{geom} = 0.1$ and $b = 5r_{mean}$, with $r_{mean} \equiv L_{box} / (N_{part})^{1/3}$ being the mean inter-particle separation, $L_{box}$ the side of the cubic simulation volume and $N_{part}$ the initial number of particles. These values are the default parameters set by \cite{garaldi_dynamic_2020}, which we also employ for comparison purposes. 

Finally, we would like to remind the reader that the gravitational framework of cosmological simulations is often Newtonian, and this is also the case in our Arepo and Gadget4 simulations. The resulting instantaneous propagation of gravitational interactions implies that the reduced resolution outside $R_{lc}$ affects, to some extent, the contents of the light cone as well. More specifically, the large-scale gravitational field needs to be preserved for DZSs to be reasonably accurate. We ensure this by setting a maximum size $L_{max}$ for nodes that can be de-refined, regulating the minimum resolution outside the light cone so that it is not too low \citep[see also][]{garaldi_dynamic_2020}. We use the default value of the maximum linear size of a node $L_{max} = 4 r_{mean}$, which roughly translates to a minimum mass resolution 64 times coarser than it would be without de-refinement.

For the sake of brevity, we only provided a somewhat general, albeit self-consistent, description of how the practical implementation of the DZS algorithm works and what are its user-defined parameters. While there are some technical differences between our implementations and the original Gadget3 one (driven by the extension to multiple, different codes and cosmological models), most of the details extensively discussed in \cite{garaldi_dynamic_2020} remain essentially unchanged. Therefore, we refer the reader to that work for further details.

\section{DZS algorithm validation} \label{sec:validation}
We now showcase our DZS implementations in terms of accuracy (i.e., the ability to reproduce cosmologically relevant observables) and performance gains on the wall-clock time taken by simulations. To this end, for a given set of initial conditions \citep[all created with the MUlti-Scale Initial Conditions tool; MUSIC;][]{hahn_multi-scale_2011}, we performed two dark-matter-only (DM-only) simulations: one with the DZS algorithm enabled (\texttt{dzs} hereafter) and one without it (\texttt{std} hereafter). We refer to such a pair as ``twin'' simulations. We employed volume and particle configurations similar to those found in \cite{llinares_shrinking_2017} and \cite{garaldi_dynamic_2020}, adapted to our range of different models and parameters, as well as to our available computational resources. Cosmological parameters were set to Planck 2018 values (\citetalias{aghanim_planck_2020} \citeyear{aghanim_planck_2020}), that is, $\Omega_{DE} = 0.6889$, $\Omega_M = 0.3111$, and $H_0 = 67.66$ km s$^{-1}$ Mpc$^{-1}$. We list our initial condition setups in Table~\ref{tab:simulations}.
\begin{table}[t]
\caption{Simulation parameters used for algorithm validation.}
\label{tab:simulations}
\centering
\begin{tabular}{c c c c c}
\hline\hline
name & $L_{box}$  & $N_{part}$ & mass res  & light cone \\
 & $[\mathrm{cMpc}/h]$ &  & $[\mathrm{M}_\odot/h]$ & $z$ range\\

\hline
   \textsc{medium} & $2048$ & $512^3$ & $5.53 \times 10^{12}$ & $[\sim0.36, 0]$ \\
   \textsc{mediumHR} & $2048$ & $1024^3$ & $6.91 \times 10^{11}$ & $[\sim0.36, 0]$ \\
   \textsc{large} & $8192$ & $512^3$ & $3.54 \times 10^{14}$ & $[\sim2.5, 0]$ \\
   \textsc{largeHR} & $8192$ & $1024^3$ & $4.42 \times 10^{13}$ & $[\sim2.5, 0]$ \\
\hline
\end{tabular}
\tablefoot{The parameters are, respectively, the side length of the simulated volume, the number of simulation particles, the resulting mass resolution, and the redshift range in which light cone output was produced. The latter is chosen to ensure that $R_{lc}(z) < L_{box}/2$ always, avoiding any box replications. Note that, at the highest considered redshifts, the DZS algorithm has already significantly modified the simulation volume.}
\end{table}
\begin{table*}[t]
\caption{Models and parameters adopted in the DZS algorithm validation.}
\label{tab:models}
\centering
\begin{tabular}{c c c c}
\hline\hline
Model name & Model ref. & Parameters & Code \\
\hline
    $\Lambda$CDM & n.n. & n.n. & Arepo \\ \\
    MG-F5 & 1 & $\Bar{f}_{R0} = -10^{-5}$ & MG-Arepo \\
    MG-F6 & 1 & $\Bar{f}_{R0} = -10^{-6}$ & MG-Arepo \\ \\
    DS-DESI & 2 & CPL, $w_0 = -0.667$, $w_a = -1.09$ & PANDA-Gadget4 \\
    DS-THAW & 3 & their Eq. 15, $w_0 = -0.667$, $w_a = -0.333$, $p=4$ & PANDA-Gadget4 \\
\hline
\end{tabular}
\tablebib{
(1)~\citet{arnold_realistic_2019}; (2) \citet{collaboration_desi_2025}; (3) \citet{lodha_extended_2025}.}
\end{table*}

In fact, our choice of DM-only setups reflects a challenge in the framework of modern cosmological simulations: the exceptionally high memory requirements needed to perform galaxy formation simulations on scales $\gtrsim 1$ cGpc$/h$. While such simulations are usually more computationally expensive their DM-only counterparts, and therefore would benefit more from DZS performance enhancements, these enhancements are effective only once the light cone enters the simulation volume. Consequently, the memory demands for starting a full-physics, large-scale simulation remain unchanged and prohibitively high for most projects (with some notable exceptions; e.g., \citealt{frontiere_exascale_2025}, or the L2p8\_m9 in \citealt{schaye_flamingo_2023}). However, these constraints do not imply that DZSs are incompatible with physics beyond gravitational interactions. Indeed, the DZS algorithm could be paired with memory-optimized initial conditions to enhance the performance of full-physics simulations. An additional complication in the full-physics case is that the new baryonic particles need to inherit more quantities from parent tree nodes than mass and phase-space localization -- for instance, conserved hydrodynamic properties -- posing the technical issue of calculating such quantities across the tree effectively. Therefore, in this work, we focus on DM-only setups and defer the implementation of a full-physics version of DZSs to future investigations.

We carried out our validation for a range of cosmological models: for MG-Arepo, we used two values for $\Bar{f}_{R0}$, i.e., $-10^{-5}$ and $-10^{-6}$. The former yields a stronger scalar field with more prominent modified gravity effects,\footnote{In fact, such a value for $\Bar{f}_{R0}$ is in tension with observations \citep[e.g.,][]{arnold_realistic_2019}; nonetheless, it is a standard parameter choice for simulations of modified gravity \citep[e.g.,][]{winther_modified_2015}, yielding clearly identifiable effects in both force enhancement and screening \citep[e.g.,][]{hagstotz_modified_2019}.} while the latter is closer to the $\Lambda$CDM case. For DS simulations with PANDA-Gadget, we employed the two $w_{DE}(z)$ parameterizations shown in Fig.~\ref{fig:wDE}, namely the \cite{collaboration_desi_2025} best-fit and \cite{lodha_extended_2025} thawing parametrization. As a reference, we also include $\Lambda$CDM runs performed with Arepo. Employed models are summarized in Table~\ref{tab:models}. For each configuration, we performed a twin pair of simulations with every setup outlined in Table~\ref{tab:simulations}. This amounts to a total of $40$ simulations, which would have been unfeasible to carry out with resolutions higher than our fiducial \textsc{mediumHR}. Based on these results, we nonetheless provide an approximate analysis for higher-resolution performance in Sect. \ref{sec:HR}.

Our validation started with an accuracy assessment on the  output of twin simulations. We chose to focus on the output that the DZS method is tailored to produce, i.e., light-cone-like output. This is because the heavy modifications introduced by the DZS method on the simulation domain (i.e., a substantial change in resolution and reduction of the total number of particles) prevent 3D power spectra or halo mass functions from being readily calculated. Note that our analysis is strongly dependent on the chosen parameters for DZSs, namely the buffer zone ($b$), opening angle ($\theta_{geom}$) and especially maximum node size to be merged ($L_{max}$), which controls the accuracy of the large-scale gravitational field. The DZS algorithm can be made arbitrarily accurate (e.g., by lowering $\theta_{geom}$ for a stricter de-refinement criterion), at the cost of a reduced performance boost. We refer to \cite{garaldi_dynamic_2020} for an extensive parameter variation analysis.

\subsection{Light cone halo mass function} \label{sec:LCHMF}
The light cone halo mass function (LCHMF) represents the number density of dark matter halos as a function of their mass, in a light-cone-like view over a certain redshift range. Arepo and Gadget4 employ the friends-of-friends (FoF) algorithm \citep[][]{davis_evolution_1985} for detecting collapsed structures from a distribution of particles. In our case, this distribution is made of the interpolated positions of particles crossing $R_{lc}$ (i.e., the main light cone output), which we ran through the FoF algorithm in postprocessing to detect halos directly on the light cone. The resulting halo masses were then binned on an arbitrary number of $21$ bins between $10^{13}$ and $10^{15}$ $\mathrm{M}_\odot/h$. Due to the limited resolution of our validation suite (Table~\ref{tab:simulations}), we computed the LCHMF only for the \textsc{mediumHR} simulations. While these simulations yield only the top end of the LCHMF, we are still able to cover $\sim 2$ orders of magnitude.
\begin{figure}[ht!]
   \resizebox{\hsize}{!}{\includegraphics{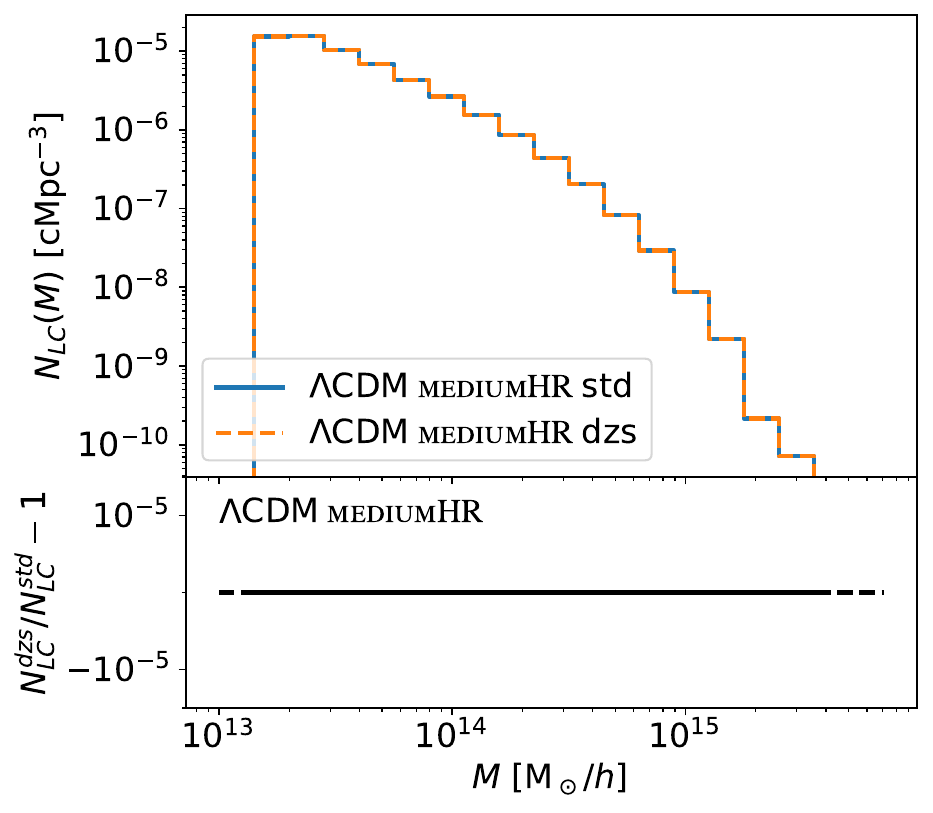}}
      \caption{Light cone halo mass function of the $\Lambda$CDM \textsc{mediumHR} twin simulations (top). The \texttt{std} solid histogram and the \texttt{dzs} dashed histogram are in excellent agreement. This is also shown by the relative difference plot between the two curves (bottom), where we arbitrarily set the $y$ axis scale.} 
         \label{fig:LCHMFlcdm}
\end{figure}
\begin{figure*}[t]
   \includegraphics[width=0.495\hsize]{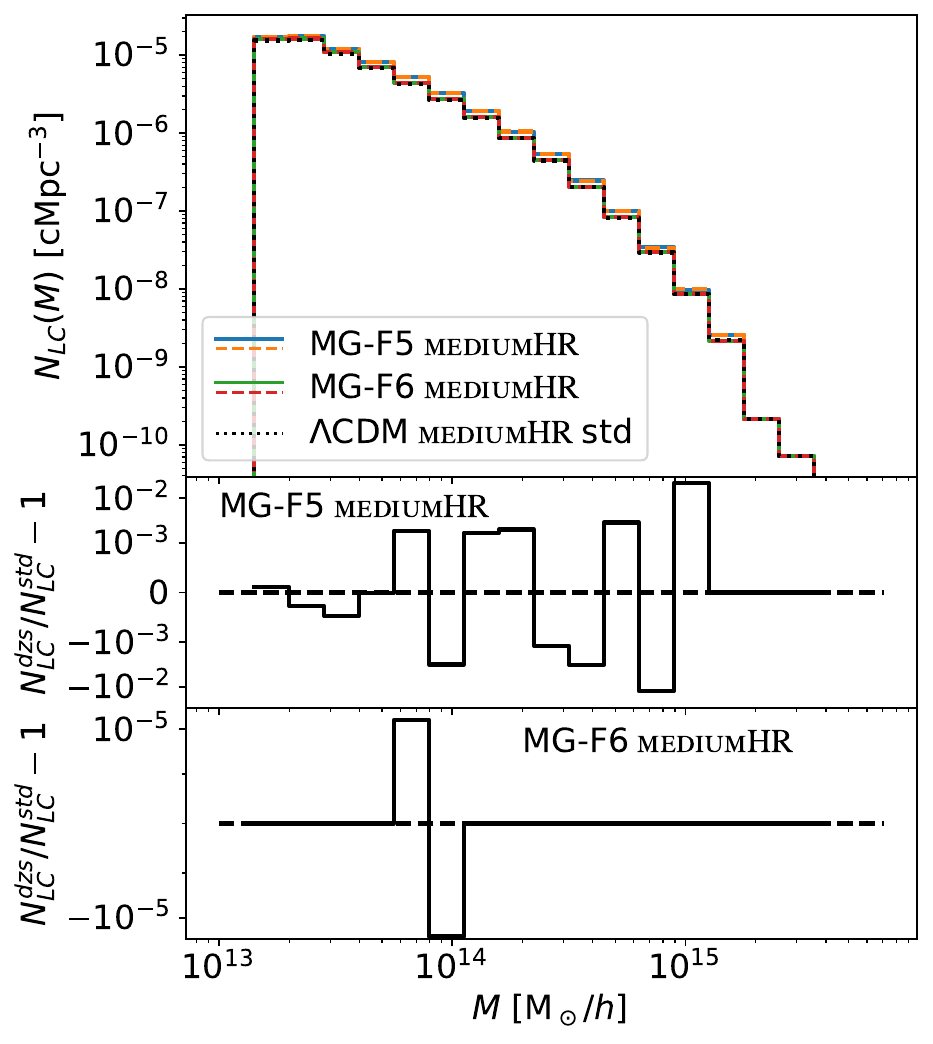}
   \includegraphics[width=0.495\hsize]{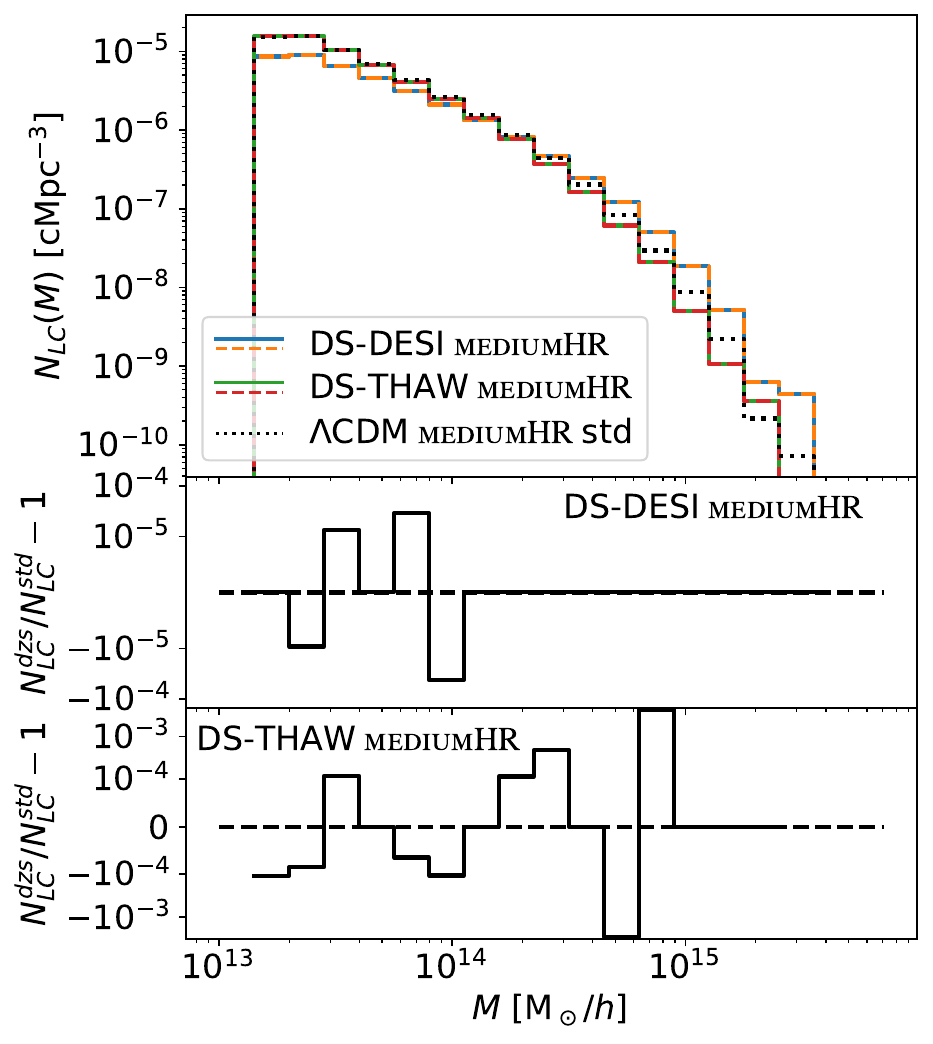}
      \caption{\textit{Left column}: Light cone halo mass function of the MG-Arepo \textsc{mediumHR} twin simulations, for the MG-F5 and MG-F6 models (\textit{top}). Solid lines refer to \texttt{std} runs, dashed lines to \texttt{dzs} ones. The $\Lambda$CDM \texttt{std} case is overlaid as a dotted line for reference. Relative difference plots are included for both MG models (\textit{middle} and \textit{bottom}), with a dashed line at $N_{LC}^{dzs}/N_{LC}^{std} - 1 = 0$ for reference. \textit{Right column}: Same but for the DS-DESI and DS-THAW \textsc{mediumHR} simulations.} 
         \label{fig:LCHMFother}
\end{figure*}
\begin{figure*}[ht!]
   \includegraphics[width=0.33\hsize]{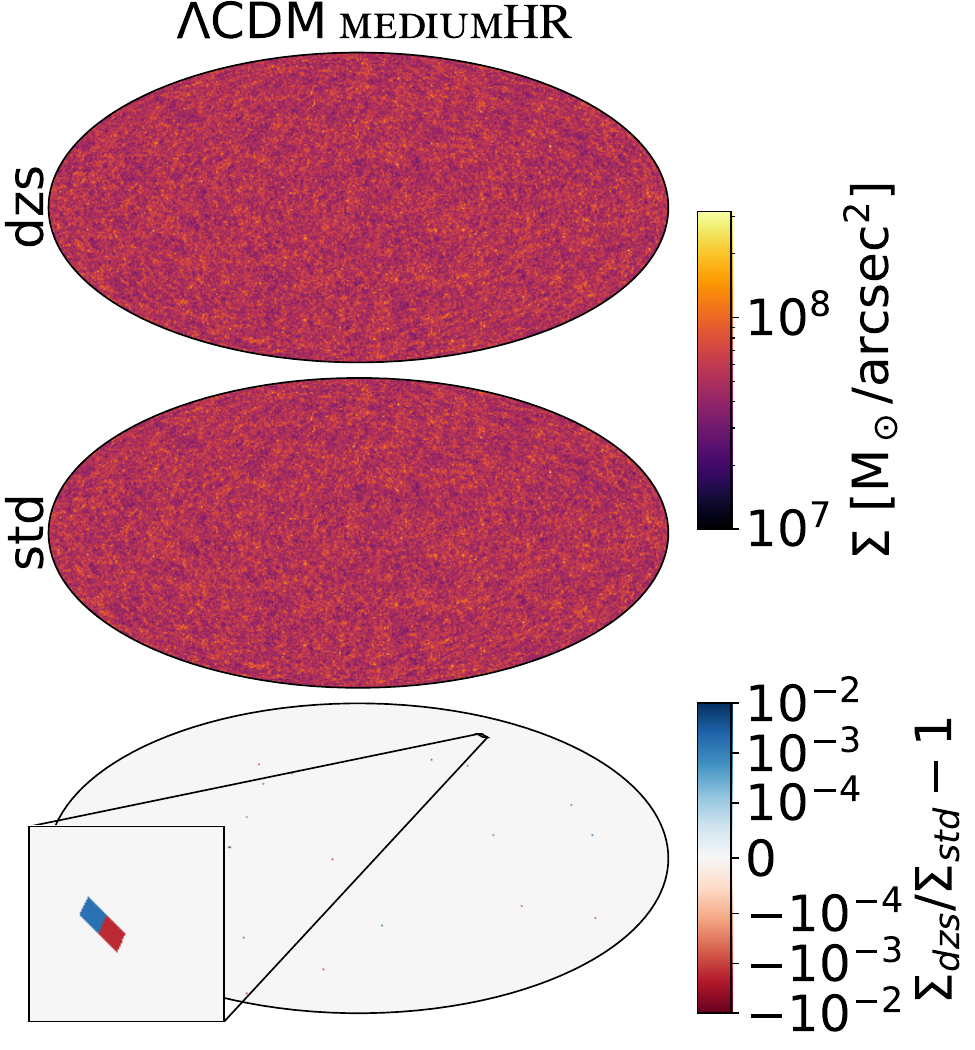}
   \includegraphics[width=0.33\hsize]{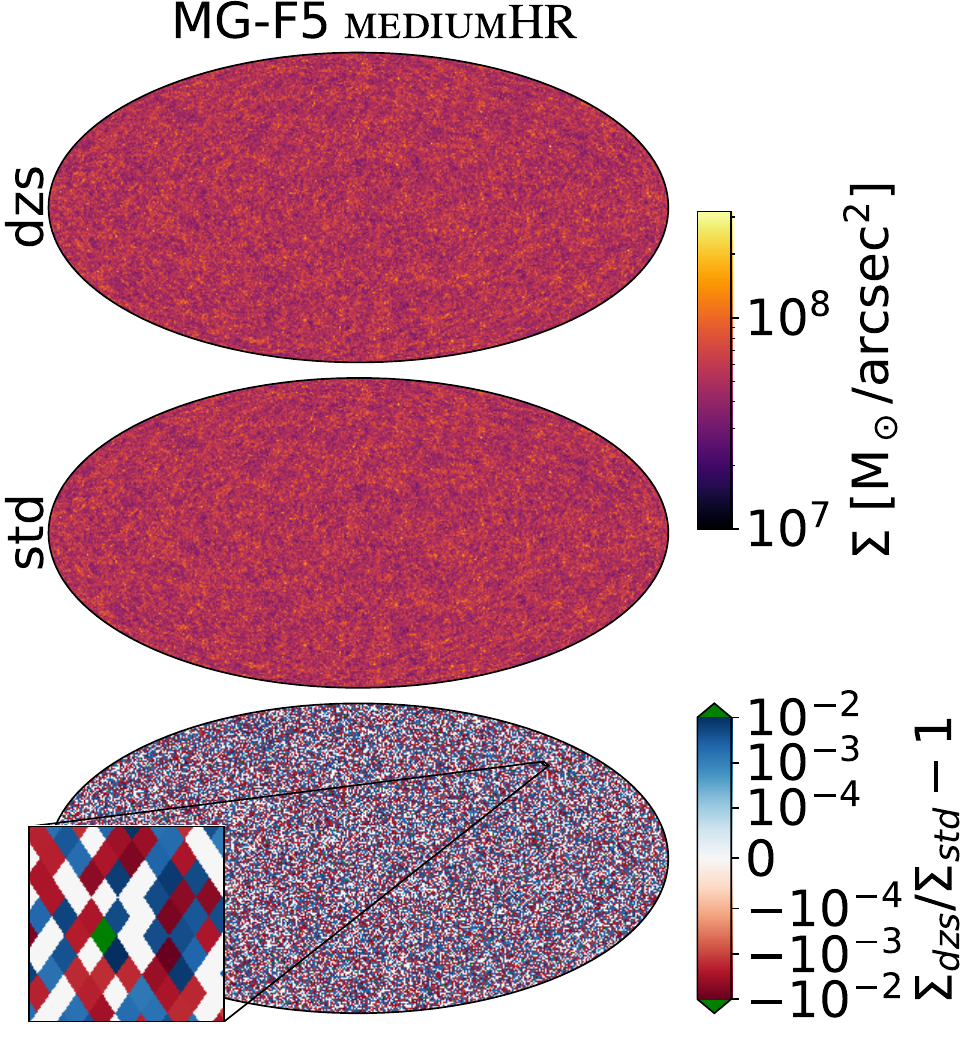}
   \includegraphics[width=0.33\hsize]{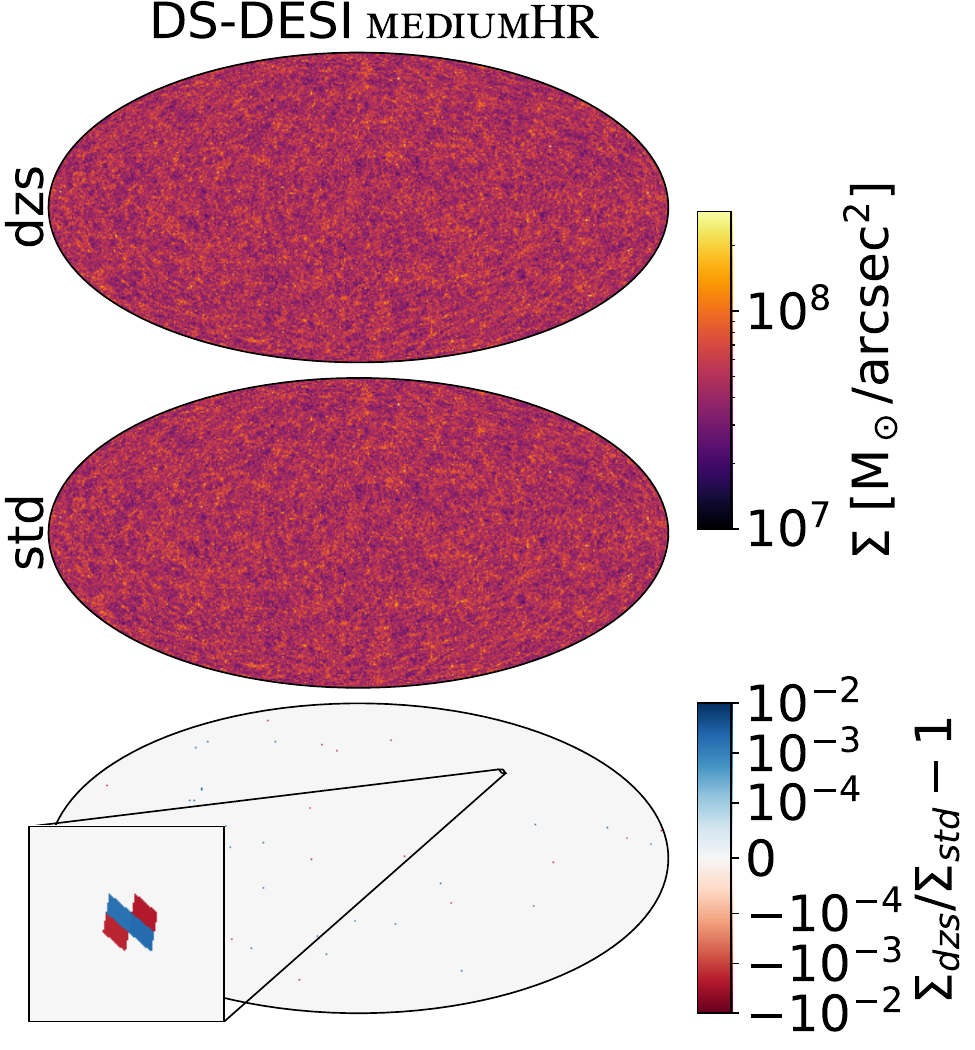}
      \caption{Maps showing the angular matter density ($\Sigma$) of the \textsc{mediumHR} twin simulations in the $\Lambda$CDM (\textit{left column}), the MG-F5 (\textit{middle column}), and the DS-DESI (\textit{right column}) cases. For each of the three models, maps for the \texttt{dzs} and \texttt{std} simulations are depicted in the top and middle panel, respectively. At the bottom, the relative differences between the \texttt{dzs} and \texttt{std} cases are shown, with the MG-F5 case employing the color green for differences above $1\%$ in absolute value.}
    \label{fig:massmaps}
\end{figure*}

We show the LCHMF for the twin \textsc{mediumHR} $\Lambda$CDM simulations in Fig.~\ref{fig:LCHMFlcdm}. The \texttt{std} simulation is reproduced exactly by the corresponding DZS. The nonstandard cases are depicted in Fig.~\ref{fig:LCHMFother}. The MG-F5 model yields relative differences of $\sim2\%$ at most between the \texttt{dzs} and \texttt{std} simulations, at the highest masses where only a handful of halos are detected (middle-left panel of Fig.~\ref{fig:LCHMFlcdm}). On the other hand, the MG-F6 model is very similar to the $\Lambda$CDM scenario, only showing $\sim 0.001\%$ differences in halos with masses around $10^{14}$ $\mathrm{M}_\odot/h$. The DS-DESI and DS-THAW models also show minimal differences, with the former always remaining below $0.01\%$ and the latter peaking at $\sim0.1\%$. With the DZS algorithm, we were also able to retrieve physical signatures of the different cosmologies. The MG-F5 model (solid blue and dashed orange lines in the top-left panel) shows an increased density of halos with respect to the reference $\Lambda$CDM case (as expected from the enhanced gravity of the $f(R)$ framework). The MG-F6 setup, where modified gravity produces weaker effects, remains closer to the reference model. The DS models show interesting LCHMF trends with respect to $\Lambda$CDM, with the DS-THAW model (solid green and dashed red lines in the top-right panel of Fig.~\ref{fig:LCHMFother}) having a lower number density of halos than the standard case, due to the factor $A(z)$ being always $>0$ (see Eq.~\ref{eq:DSdrag}). For the DS-DESI case, $A(z) < 0$ in the phantom regime, resulting in a shift of collapsed halos toward high masses. We conclude that DZSs reproduce the LCHMF of our highest resolution setup with high accuracy in all of the examined nonstandard cosmologies, with the highest relative difference being $\sim 2\%$ in the MG-F5 model at the largest halo masses ($\gtrsim 5 \times 10^{14}$ $\mathrm{M}_\odot/h$). Note that a larger halo statistics might very well improve this result.

While \cite{garaldi_dynamic_2020} also tested the accuracy of DZSs in Gadget3 through the LCHMF, a direct comparison is made difficult by the different $N_{part}$ employed (they used $N_{part} = 2048^3$), LCHMF building technique (a piecewise-constant approximation) and redshift range ($z \in [\sim 0.68,0]$). Nonetheless, we note that they achieved $\sim 0.2\%$ accuracy, comparable to our DS-THAW result and better than the MG-F5 one. This is consistent with the results that will be discussed in Sect. \ref{sec:hpix}, which show that MG setups tend to yield a slightly lower accuracy with respect to $\Lambda$CDM and DS models, independently of the code employed.

\subsection{Sky-projected light cone and angular power spectrum} \label{sec:hpix}
In addition to the 3D distribution of light-cone-crossing particles employed in the LCHMF calculation, the on-the-fly light cone output capabilities also allow us to save a sky-projected HEALPix \citep[][]{gorski_healpix_2005} pixelization of those particles. We compared the resulting angular matter density maps to assess pixel-scale relative differences, and examined angular power spectra of matter clustering. We created density maps for the \textsc{mediumHR} simulation, which has the highest resolution in the validation suite, and computed power spectra for the \textsc{mediumHR}, \textsc{medium} and \textsc{largeHR} simulations. We used a value $\mathrm{NSIDE}=1024$ for our HEALPix output,\footnote{In a HEALPix pixelization, NSIDE represents the number of pixels per side, and relates to the total number of pixels in a map NPIX as $\mathrm{NPIX} = 12 \times \mathrm{NSIDE}^2$.} resulting in a maximum multipole moment $l \simeq 3000$ in the power spectra. We averaged the value of each moment across 80 logarithmically spaced bins. Angular matter density maps were down-sampled to $\mathrm{NSIDE}=256$ for visual clarity.

Figure~\ref{fig:massmaps} shows the angular matter density (labeled as $\Sigma$) maps of the \textsc{mediumHR} twin simulations, run with the $\Lambda$CDM, MG-F5, and DS-DESI models. The $\Lambda$CDM and DS-DESI maps are almost identical in the \texttt{std} and \texttt{dzs} case, having only a few ``outlier'' pixels with nonzero relative difference of $1\%$ at most in absolute value. On the other hand, the MG-F5 result is noticeably different. This is due to the MG contribution being calculated at node level (that is, on each cell of the multi-grid solver), and then mapped to the particles in each node. The extremely small positional displacements induced by DZSs in the particle distribution (even of the order of one-billionth of the spatial resolution) can move particles from one node to another -- or  prevent the creation of finer grid cells altogether -- and yield different MG contributions at particle level. The fact that these differences arise within the light cone leads to faster-propagating and more significant changes in the output, especially when the light cone crossing of individual particles is considered for such output. Therefore, the multi-grid solver has effects on the accuracy of MG models -- especially MG-F5 -- which are not present in our $\Lambda$CDM and DS models, nor in \cite{garaldi_dynamic_2020}. Nonetheless, as shown in Fig.~\ref{fig:massmaps}, the accuracy is still at percent level at worst, and only a small fraction of pixels ($0.9\%$ of the total, highlighted in green) yields differences greater than $1\%$ in absolute value.

\begin{figure}[t]
   \resizebox{\hsize}{!}{\includegraphics{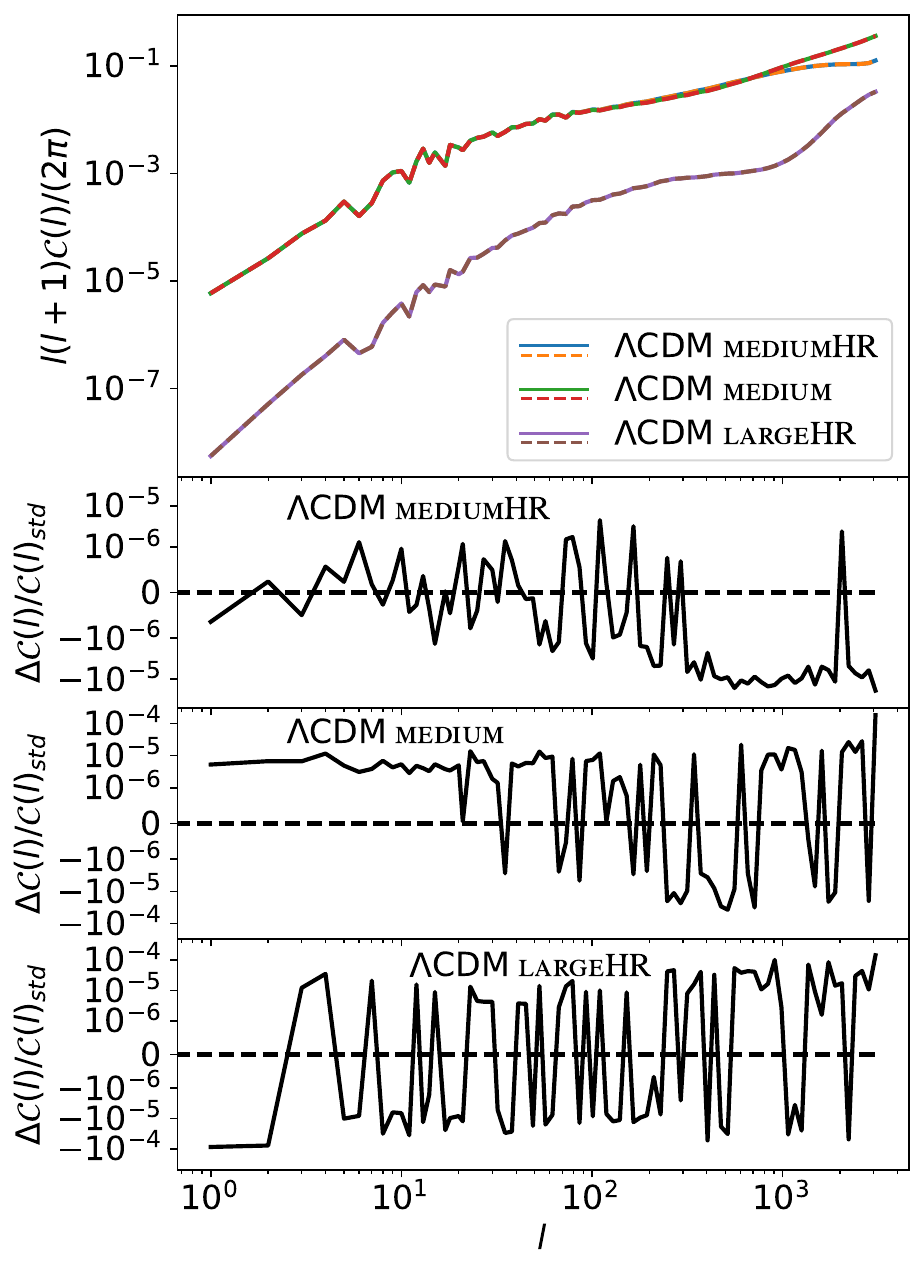}}
      \caption{\textit{Top panel}: Angular power spectrum ($\mathcal{C}(l)$) of $\Lambda$CDM \textsc{mediumHR}, \textsc{medium,} and \textsc{largeHR} runs. The power is lower in the latter pair of simulations due to the larger output redshift range (see Table~\ref{tab:simulations}). Solid lines refer to \texttt{std} runs, dashed lines to \texttt{dzs} ones. \textit{Bottom panels}: Relative differences between the \texttt{dzs} and \texttt{std} simulations for the different run types discussed above, as indicated in each panel. The dashed line is located at $\Delta \mathcal{C}(l)=0$ as a reference.} 
         \label{fig:pslcdm}
\end{figure}
\begin{figure*}[ht!]
   \includegraphics[width=0.495\hsize]{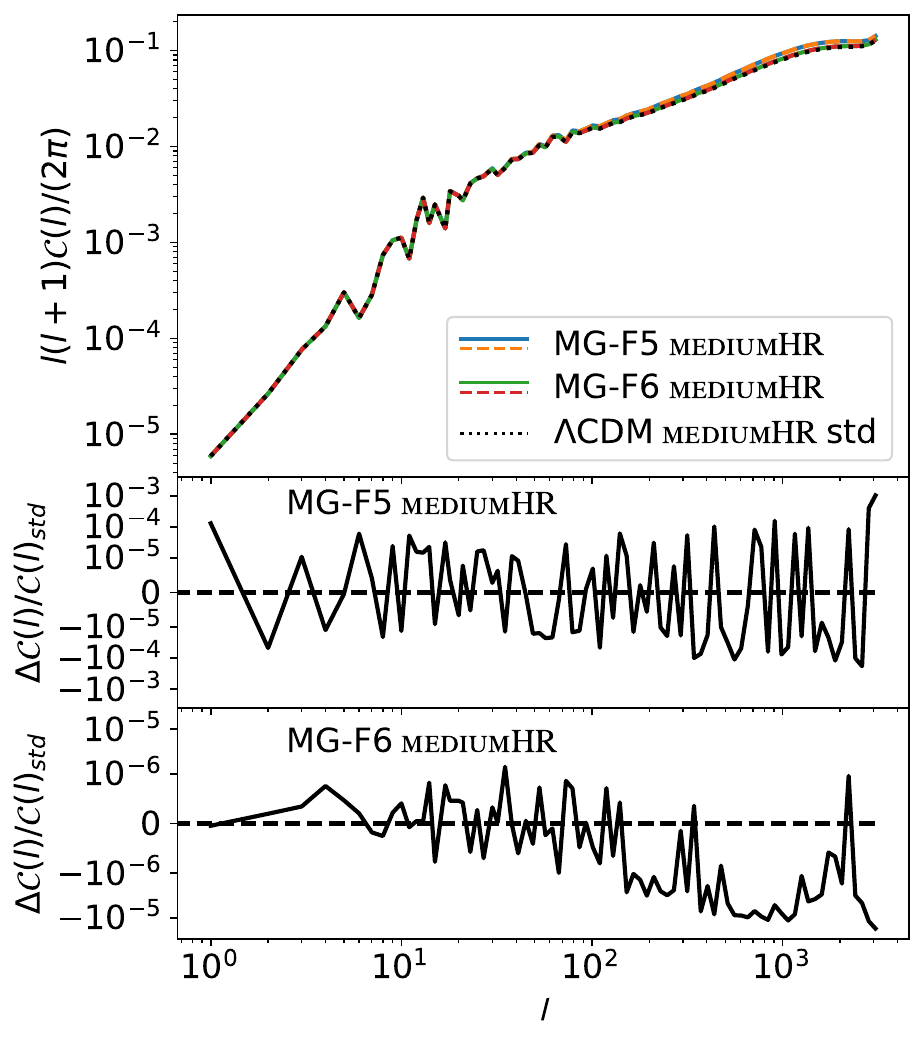}
   \includegraphics[width=0.495\hsize]{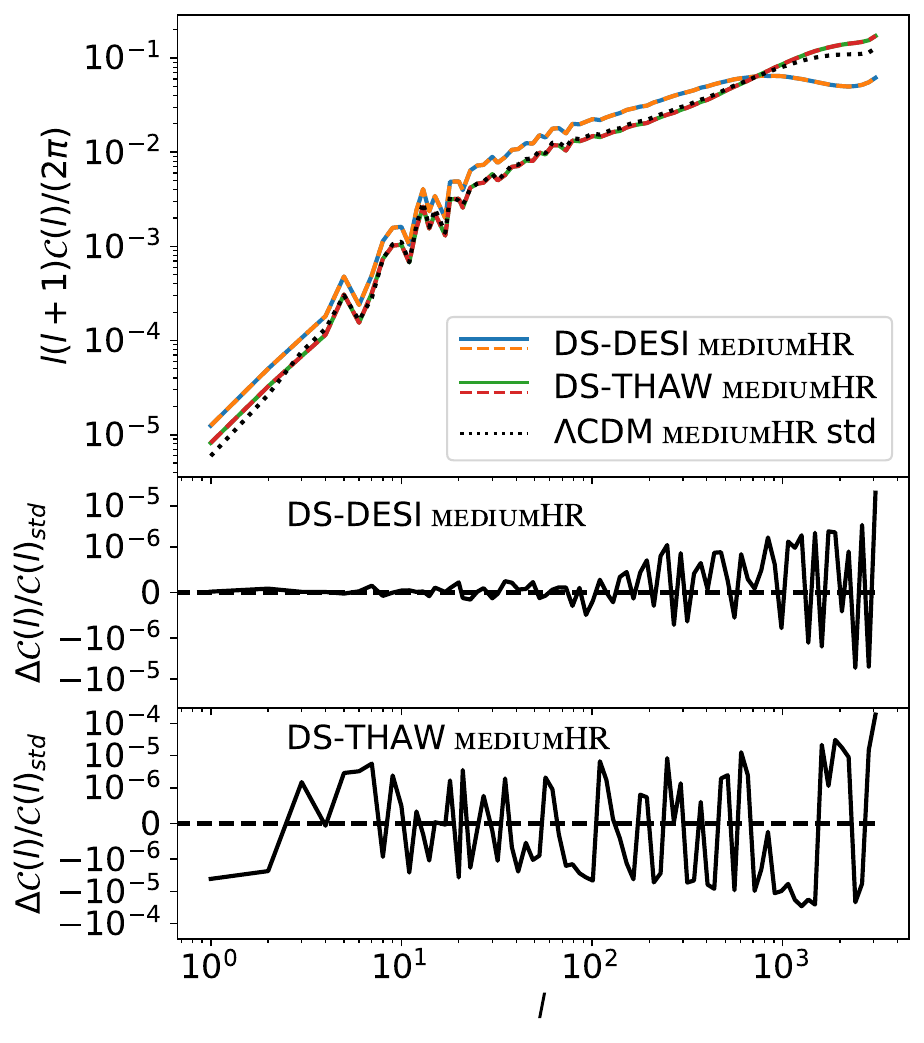}
      \caption{\textit{Left column}: Angular power spectrum ($\mathcal{C}(l)$) of the MG-Arepo \textsc{mediumHR} simulations, for the MG-F5 and MG-F6 models (\textit{top}). Solid lines refer to \texttt{std} runs, dashed lines to \texttt{dzs} ones. The $\Lambda$CDM \texttt{std} case (dotted line) is overlaid as a reference. Relative difference plots are included for both MG models (\textit{middle} and \textit{bottom}, as labeled), with a dashed line indicating  $\Delta \mathcal{C}(l)=0$ for reference. \textit{Right column}: Same but for the DS-DESI and DS-THAW \textsc{mediumHR} simulations.} 
         \label{fig:psmediumHR}
\end{figure*}
\begin{figure}[ht!]
   \resizebox{\hsize}{!}{\includegraphics{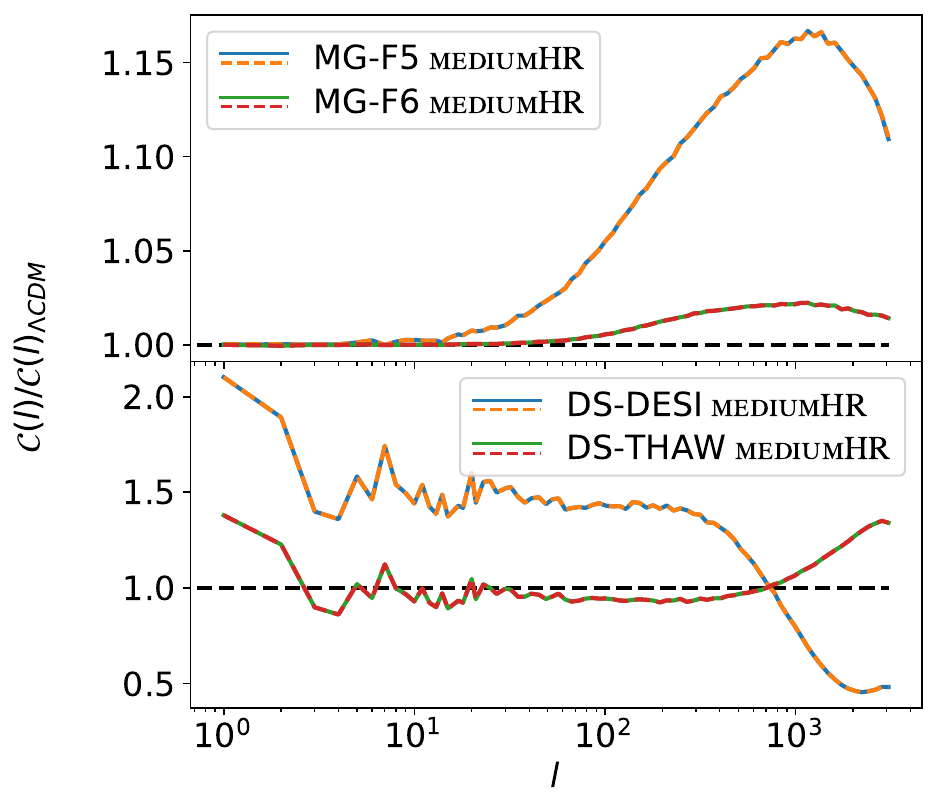}}
      \caption{Power spectrum boosts, i.e., the ratio $\mathcal{C}(l)/\mathcal{C}(l)_{\Lambda\mathrm{CDM}}$, of the \textsc{mediumHR} simulations, for MG models (\textit{top}) and DS ones (\textit{bottom}). Dashed lines at unit boost are shown for reference.} 
         \label{fig:psboost}
\end{figure}

Figure~\ref{fig:pslcdm} depicts the matter power spectrum for the $\Lambda$CDM \textsc{mediumHR}, \textsc{medium}, and \textsc{largeHR} twin simulations (due to their low resolution, the \textsc{large} simulations are not included and will only be used for performance estimation). Pairs of solid lines (\texttt{std} simulations) and dashed lines (\texttt{dzs} simulations) refer to different setups. Due to the broader redshift range of the light cone output in the \textsc{largeHR} runs, which encompasses structures at higher $z$ than the \textsc{medium} and \textsc{mediumHR} boxes (akin to a ``deeper survey''), the density contrast is less pronounced and the power is significantly lower. As for the \textsc{medium} and \textsc{mediumHR} simulations, the same box at a different resolution yields visually identical results with the exception of the smallest scales (largest $l$) sampled, where the lower resolution shows a higher power. The accuracy of the DZS algorithm is very high in the $\Lambda$CDM framework: solid and dashed lines look exactly the same to the naked eye, and the relative difference plots (bottom panels in Fig.~\ref{fig:pslcdm}) confirm this trend. Differences sit at $\sim 0.01\%$ at worst in the \textsc{largeHR} and \textsc{medium} setups. Note that the latter is outperformed in accuracy by its higher-resolution counterpart: this trend is very promising for the use of the DZS algorithm in state-of-the-art, high-resolution cosmological simulations. Power spectrum differences for the \textsc{largeHR} were also presented in \cite{garaldi_dynamic_2020}, albeit in a smaller redshift and $l$ range ($z \in [1.5,0]$, $l \in [0,200]$). In the same range of $l$, we register $\sim 0.01\%$ accuracy, whereas Gadget3 yielded $\sim 0.1\%$. This discrepancy is expected because of the broader redshift range analyzed in our simulations, over which differences tend to average out more effectively. Finally, for the sake of completeness, we note that the \textsc{mediumHR} runs show seemingly nonrandom differences at small scales, with the \texttt{dzs} simulation having slightly ($\sim 0.001\%$) less power. Since this difference is so small, we did not investigate it further.

The beyond-$\Lambda$CDM cases are shown in Fig.~\ref{fig:psmediumHR} for the \textsc{mediumHR} simulations. As it was the case in the LCHMF comparison, the MG-F5 \textsc{mediumHR} simulations (solid blue and dashed orange lines in the top-left panel of Fig.~\ref{fig:psmediumHR}) yield the worst results among our validation setups (as shown by the relative differences in the middle-left panel of the same figure). Nonetheless, relative differences remain around $0.01\%$ for most values of $l$, only rising to $0.1\%$ at the lower end of the sampled scales, corresponding to $l \gtrsim 10^3 $. The power spectrum itself also shows a boost with respect to the reference $\Lambda$CDM case, in accordance with the LCHMF (Fig.~\ref{fig:LCHMFother}) and the theoretical $f(R)$ framework. The MG-F6 \textsc{mediumHR} simulations, instead, are more similar to the standard setup both in power spectrum boost (solid green and dashed red line in the top-left panel of Fig.~\ref{fig:psmediumHR}) and relative differences (bottom-left panel of the same figure). The PANDA-Gadget4 \textsc{mediumHR} simulations (top-right panel of Fig.~\ref{fig:psmediumHR}), are also consistent with their LCHMF: the DS-DESI model is boosted at larger scales and suppressed at smaller scales with respect to $\Lambda$CDM, whereas the thawing model (DS-THAW) has slightly less power for most values of $l$. The comparison between \texttt{std} and \texttt{dzs} runs (middle-right and bottom-right panel of Fig.~\ref{fig:psmediumHR}) yields extremely small, visually random differences. Specifically, these remain below $\sim 0.001\%$, except for $l \gtrsim 10^3$ in the DS-THAW model, for which differences increase by approximately one order of magnitude.

Figures~\ref{fig:psmedium} and \ref{fig:pslargeHR} contain the power spectra of the nonstandard \textsc{medium} and \textsc{largeHR} simulations. Here, we remark that the DZS-induced differences are very contained in those setups as well. Furthermore, a comparison between the \textsc{medium} and the \textsc{mediumHR} runs shows that, unlike the standard setup of Fig.~\ref{fig:pslcdm}, the differences in some models (namely MG-F5 and DS-THAW) do not improve with resolution; we note, however, that they do not get worse either. We expect this behavior to continue in the high-resolution, large-scale scenarios that the DZS algorithm is tailored for. More generally, we note that all of the relative differences depicted in Figs.~\ref{fig:pslcdm}, \ref{fig:psmediumHR}, \ref{fig:psmedium} and \ref{fig:pslargeHR} are well within the accuracy forecasts of modern (i.e., ``Stage-IV'') surveys. Such forecasts report a $1\%$ accuracy requirement on the 3D power spectrum $P(k)$ \citep[e.g.,][]{schneider_percent_2016, taylor_shear_2018}. While error mapping between $P(k)$ and the sky-projected spectrum depends on the cosmological model and examined redshift range, we do not expect this mapping procedure to significantly affect the primary accuracy requirement on $P(k)$.

Finally, we highlighted the physical differences between the $\Lambda$CDM power spectrum and those of other models through their ratio (i.e., the so-called boost), shown in Fig.~\ref{fig:psboost} for the \textsc{mediumHR} simulations. As hinted earlier, modified gravity setups (top panel) show a boost up to $15\%$ in the MG-F5 model (at $l \sim 10^3$), with milder effects, albeit still of a few percent, in the MG-F6 case. Dark scattering models (bottom panels) show the opposite behavior that we already discussed in Sect. \ref{sec:LCHMF}, due to the different signs of $A(z)$ (albeit with milder effects in the DS-THAW case). We omit the relative differences for boosts specifically, since they are essentially comparable to those of the beyond-$\Lambda$CDM models (Fig.~\ref{fig:psmediumHR}). Therefore, such differences remain also well within observational forecasts for Stage-IV surveys, as discussed above. Furthermore, \citetalias{collaboration_euclid_2024-1} (\citeyear{collaboration_euclid_2024-1}) directly compares boosts from beyond-$\Lambda$CDM numerical codes, finding percent-level agreement, albeit again at the level of the 3D power spectrum.

\subsection{Weak lensing convergence maps} \label{sec:lensing}
We used the \texttt{dorian} Python package\footnote{\texttt{https://gitlab.mpcdf.mpg.de/fferlito/dorian}.} \citep{ferlito_ray-tracing_2024} to produce weak lensing convergence maps and power spectra from the sky-projected light cone matter density maps. The code performs a weak gravitational lensing analysis through ray-tracing of a source at a certain redshift. Due to the resolution of our simulations, we are limited to $\mathrm{NSIDE}=1024$. While this is a low value for the standards of modern weak lensing analysis, we are still able to compute the convergence power spectrum and assess its accuracy with the DZS algorithm. We do so for the \textsc{mediumHR} simulations, across two of our beyond-$\Lambda$CDM models. We anticipate that using a higher value of NSIDE, i.e., running a higher-resolution simulation, could yield improved (or, at worst, similar) relative differences in convergence power spectra with respect to our $\mathrm{NSIDE}=1024$ results. This follows from the behavior observed in the matter power spectra at different resolutions (Figs.~\ref{fig:psmediumHR} and \ref{fig:psmedium}).
\begin{figure}[ht!]
   \resizebox{\hsize}{!}{\includegraphics{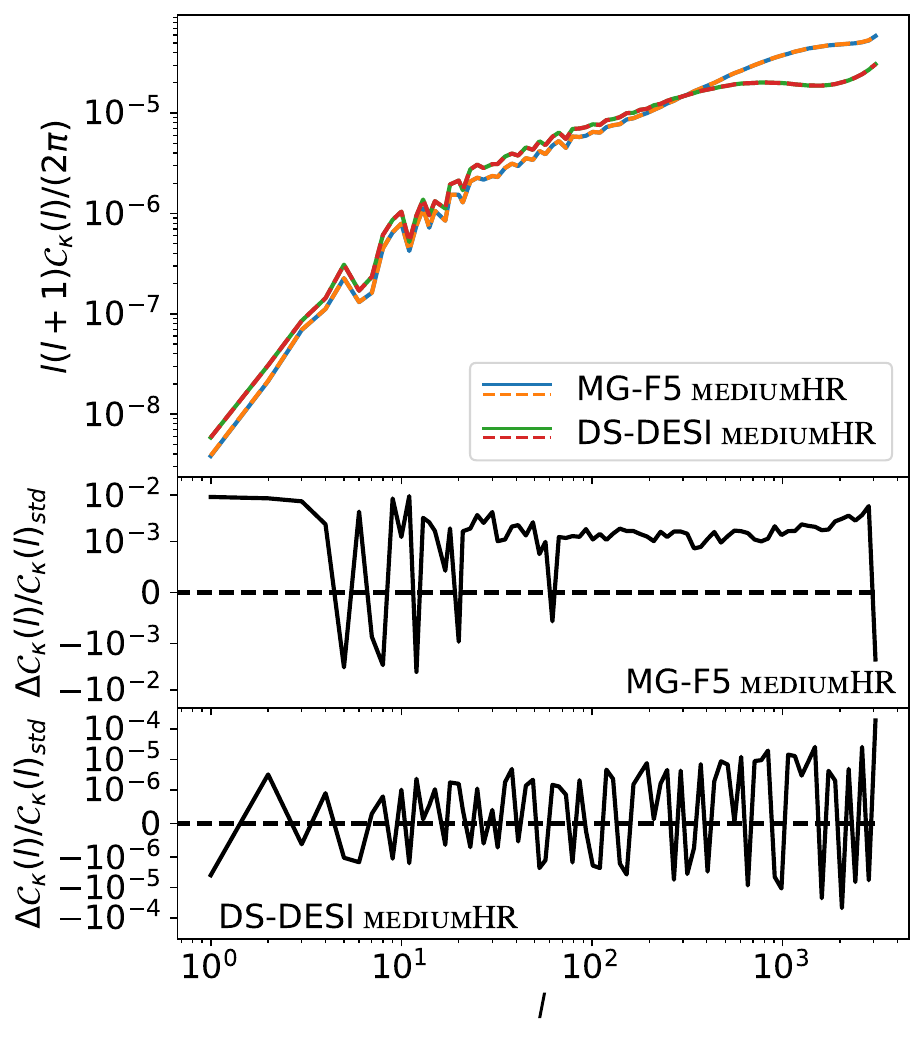}}
      \caption{Weak lensing convergence power spectrum ($\mathcal{C}_\kappa(l)$) of the \textsc{mediumHR} simulations, for the MG-F5 and DS-DESI models (\textit{top}). Solid lines refer to \texttt{std} runs, dashed lines to \texttt{dzs} ones. Relative difference plots are included for both MG models (\textit{middle} and \textit{bottom}), with a dashed line at $\Delta \mathcal{C}_\kappa(l)=0$ for reference.} 
         \label{fig:pskappamediumHR}
\end{figure}

Figure~\ref{fig:pskappamediumHR} shows the results of our analysis for the MG-F5 and DS-DESI models. We note a $\leq1\%$ accuracy across all the range of $l$ for the former model and a $0.01\%$ accuracy at worst for the latter one. We note that these results are orders of magnitude higher (in absolute value) than those of the corresponding matter power spectra (Fig.~\ref{fig:psmediumHR}). In fact, the mass maps that give rise to the matter power spectrum are built through a pixel-wise sum of several individual ``shells'' ($30$, for the \textsc{mediumHR} volume), each spanning a small redshift range to collectively cover $z \in [\sim0.36, 0]$. This tends to smooth out differences in individual shells. The \texttt{dorian} code, on the other hand, goes through each shell sequentially, resulting in more significant differences. Nonetheless, we remark that the accuracy of the DZS method is high also for the weak lensing convergence power spectra, as it also fits within the $1\%$ observational requirement discussed in Sect. \ref{sec:hpix}, and especially so in the DS-DESI case. 

\subsection{Run time performance} \label{sec:performance}
We verified that the DZS algorithm is accurate in reproducing light-cone-like output. Now, we quantify how much computational time we are able to save in our validation suite across the different cosmologies examined. Table~\ref{tab:times} shows the percentage of time saved by a \texttt{dzs} simulation with respect to the corresponding \texttt{std} run. The speedup, here defined as the total time taken by the \texttt{std} run divided by the \texttt{dzs} one, is shown in brackets. As expected, a larger volume yields higher computational time savings in every cosmological model. Moreover, a comparison between the \textsc{large} and \textsc{largeHR} simulations reveals that an increased resolution results in a larger time saving. We expect this trend to persist when pushing for higher, state-of-the-art resolutions, enabling massive performance gains (see Sect. \ref{sec:HR} for an approximate analysis on this aspect). This trend also appears in the smaller volumes simulated, albeit with the notable exceptions of the $\Lambda$CDM and MG-F6 models where the time saving in the \textsc{mediumHR} setup is $\approx 1\%$ lower than that of the \textsc{medium} one. While we ran every simulation in our testing suite on computational units with identical hardware specifications, fluctuations among different units (and on the same unit at different times) are always present. Therefore, we ascribe the somewhat inconsistent trend in the smaller volume to machine fluctuations coupled with a relatively low impact of the DZS algorithm, which only operates at redshift $\lesssim 0.69$.

While the $\Lambda$CDM trends across different setups are consistent with those reported by \cite{garaldi_dynamic_2020}, notable differences arise in other cosmological models. Specifically, the best-performing model in every setup is MG-F5, for which we saved an impressive $\sim53\%$ of the computing time in the \textsc{largeHR} simulations. This is expected, given the computationally demanding multi-grid technique employed to calculate MG effects, coupled with the high strength of the scalar field responsible for them. The performance of the multi-grid technique depends on the number of particles of a simulation, as well as their degree of clustering; therefore, by drastically reducing both the number of particles and their resolution, the DZS technique is able to yield a high performance gain in MG simulations. The presence of the same multi-grid technique, albeit with lower field strength, brings the MG-F6 to a close second place in terms of performance gain. Notably, as hinted above, the parameters on which the DZS algorithm depends can be tweaked to achieve better performance at the cost of accuracy, and vice versa. In principle, the MG setups could therefore be made more accurate -- i.e., on par with the $\Lambda$CDM and DS-DESI models in Fig.~\ref{fig:massmaps} --  
by reducing their performance gain, which would then also be on par with $\Lambda$CDM and DS-DESI results. Nonetheless, we note that the same DZS parameter choices lead to an inherently higher performance gain in MG models.

On the other hand, the DS setups show slightly lower performance gains than all of the other models. We speculate that this could be due to Gadget4 being harder to optimize than Arepo on some systems, somewhat hindering the achievable performance. In any case, the performance gain is up to $43\%$ in all of the \textsc{large} and \textsc{largeHR} simulations, and significant ($\lesssim 18\%$) also in the smaller setups (especially in MG models). We also note that DZS-related operations themselves (tree walk and particle addition/elimination) do not add any significant overhead, as the wall-clock time taken by them is always of the order $\sim 0.1\%$ of the total simulation time.

In addition to this run-time performance gain analysis, Appendix \ref{sec:wl} details the workload balance of our validation suite, to see how DZS-related modifications to particle number and resolution affect the workload distribution between parallel tasks. In the case of imbalances in this distribution, further performance improvements could be achieved through dedicated optimization techniques.

\begin{table*}[ht!]
\caption{Percentage of run time saved by \texttt{dzs} relative to \texttt{std} runs, with the corresponding speedup (in brackets) for every examined combination of initial conditions (columns) and cosmological model (rows).}
\label{tab:times}
\centering
\begin{tabular}{c | c c c c}
\hline\hline
 & \textsc{medium} & \textsc{mediumHR} & \textsc{large} & \textsc{largeHR}\\
\hline
   $\Lambda$CDM & $18.39\%$ ($1.23$) & $17.76\%$ ($1.22$) & $33.46\%$ ($1.50$) & $49.09\%$ ($1.96$) \\
   MG-F5 & $25.99\%$ ($1.35$) & $27.14\%$ ($1.37$) & $36.99\%$ ($1.59$) & $52.79\%$ ($2.12$) \\
   MG-F6 & $23.93\%$ ($1.31$) & $22.47\%$ ($1.29$) & $36.33\%$ ($1.57$) & $51.36\%$ ($2.06$) \\
   DS-DESI & $8.68\%$ ($1.09$) & $18.17\%$ ($1.22$) & $29.14\%$ ($1.41$) & $43.24\%$ ($1.76$)\\
   DS-THAW & $11.95\%$ ($1.14$) & $17.53\%$ ($1.21$) & $25.73\%$ ($1.35$) & $42.79\%$ ($1.75$)\\
\hline
\end{tabular}
\end{table*}

\section{High-resolution performance gain estimation} \label{sec:HR}

\begin{figure}[b]
   \resizebox{\hsize}{!}{\includegraphics{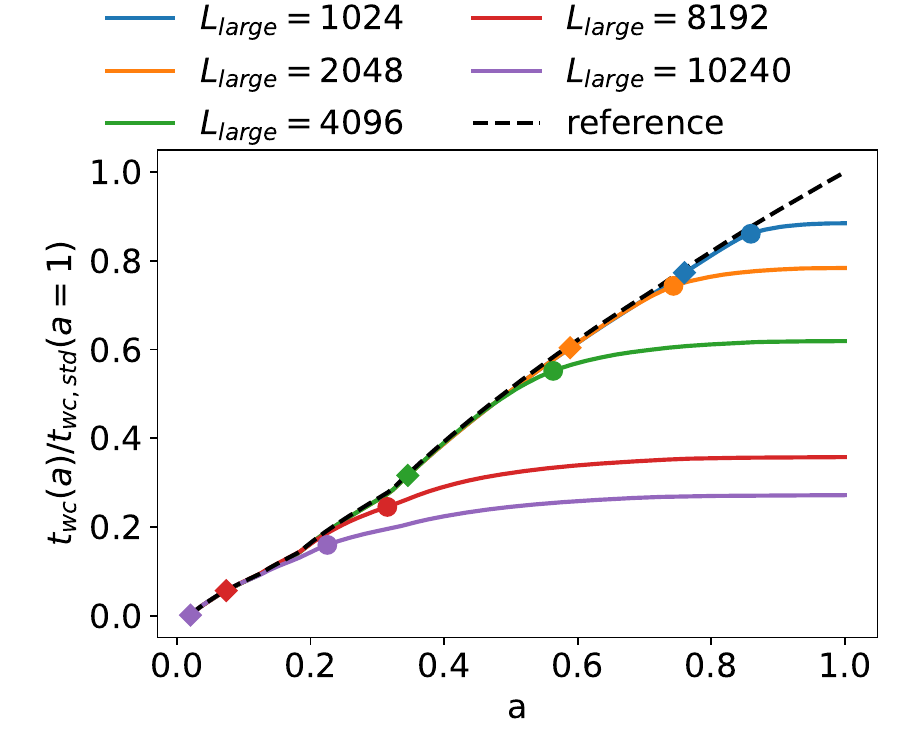}}
      \caption{Cumulative wall-clock times ($t_{wc}(a)$) of each rescaled \texttt{dzs} run of the $L_{box} = 100$ cMpc$/h$, $N_{part} = 512^3$ setup as a function of the scale factor ($a$). The times are normalized to the final (total wall-clock) time of a \texttt{std} run with the same setup. The latter simulation is also included as a dashed black line for reference. The diamond symbol marks the time at which the rescaled light cone enters the simulation volume and the DZS algorithm starts merging particles. The circular symbol shows when the number of particles becomes lower than half of the initial value as a result of the merging operation. The values of $L_{large}$ are in cMpc$/h$.}
         \label{fig:hires}
\end{figure}
While the testing suite employed so far has proved that the DZS algorithm is able to boost the performance of large-scale simulations without significant accuracy losses, such a suite does not represent the state-of-the-art simulations that our method is targeted at. As mentioned previously, we needed computationally inexpensive simulations to thoroughly test our chosen cosmological models. Here, we provide an approximate estimate of the DZS performance gain at a more realistic resolution for production runs. Note that, due to the limited computational resources available, we carried out this analysis only for $\Lambda$CDM cosmology. We expect more expensive simulation setups (such as those performed with MG-Arepo) to yield a higher performance gain, as discussed in Sect. \ref{sec:performance}. We also point out that a similar analysis was performed by \cite{garaldi_dynamic_2020}, by means of a simple linear approximation on the wall-clock time evolution of DZSs. Here, we chose a more direct strategy, which is expected to yield more accurate results.

Our starting element in this analysis was a new simulation setup with $L_{box} = 100$ cMpc$/h$ and $N_{part} = 512^3$, which corresponds to a mass resolution of $6.43 \times 10^{8}$ M$_\odot/h$ with the cosmological parameters adopted in this work. This resolution is comparable, for example, with the MillenniumTNG and \textit{Euclid} flagship simulations, therefore we consider it representative of state-of-the-art, large-scale simulations. Since the small $L_{box}$ value does not normally yield a significant DZS-related performance gain (as the light cone would enter the volume at a redshift close to zero), we rescaled the real light cone by a user-defined factor that mimics the evolution of $R_{lc}(z)$ in larger boxes with side $L_{large}$, such that $R_{lc,rescaled}(z) = R_{lc}(z)L_{box}/L_{large}$. We chose different values for $L_{large}$, namely $1024$, $2048$, $4096$, $8192$, and $10240$ cMpc$/h$. We then compared the run times of \texttt{dzs} simulations employing $R_{lc,rescaled}$ with a reference \texttt{std} run. While such an approach yields a performance gain estimate for large-scale, high-resolution simulations, the use of a small $L_{box}$ implies the absence of large-scale density fluctuations that significantly enhance the gravitational potential and, consequently, nonlinearity and computational cost. On the other hand, a real state-of-the-art setup would require a massively parallel environment with tens (or hundreds) of thousands of simultaneous tasks, where workload imbalances arise as a result of DZS-related operations (see Appendix \ref{sec:wl}). Nonetheless, we believe this to be an interesting and relatively inexpensive approach to assess the capabilities of our method in realistic setups.

Figure~\ref{fig:hires} shows the cumulative wall-clock time, $t_{wc}(a)$, taken by each of the examined values of $L_{large}$, as well as by the reference run itself, as a function of the scale factor, $a$. The values are normalized to the redshift-zero value of the reference simulation $t_{wc,std}(a=1)$. First of all, we note that each curve (except the ``reference'' one) flattens after the DZS algorithm has significantly reduced the number of particles in the simulation. Interestingly, a noticeable flattening is present also in the smallest $L_{large} = 1024$ cMpc$/h$, corresponding to a run time saving of $12\%$. As expected, larger volumes show increasingly better results, from $22\%$ with $L_{large} = 2048$ cMpc$/h$ to $64\%$ and $73\%$ in the largest re-scalings of $L_{large} = 8192$ cMpc$/h$ and $10240$ cMpc$/h$, respectively. The latter case ideally translates into a flagship-like simulation, with unique light cone output (i.e., without volume replications) available up to redshift $\sim 4$, completed in just $1/4$ of the time that it would normally take. Moreover, we note that our intermediate rescaled volume $L_{large} = 4096$ cMpc$/h$ is only slightly larger than that of the \textit{Euclid} flagships, and yields a run time saving of $38\%$. This suggests that a similar saving could be achieved by running those simulations with the DZS algorithm. Overall, these results imply that the DZS method can bring major performance improvements to state-of-the-art, large-scale simulations, also enabling multiple simulations (with different cosmologies) in the time span normally taken by a single run.

Furthermore, the $2048$ and $8192$ cMpc$/h$ re-scalings enable a direct comparison with the $\Lambda$CDM results in Table~\ref{tab:times}: both the \textsc{medium} and \textsc{mediumHR} simulations are slightly outperformed by $L_{large} = 2048$ cMpc$/h$. More noticeably, rescaling the high-resolution simulation to $8192$ cMpc$/h$ saves $30\%$ more time than in the \textsc{large} case, and $15\%$ more time than in the \textsc{largeHR}. This agrees with the trend originally observed in Sect. \ref{sec:performance}; namely, increasing the resolution on a fixed volume also increases the run time saved thanks to the DZS method. Lastly, we note that rescaling to $L_{large} = 10240$ cMpc$/h$ mimics a volume so large that it is not fully contained in the light cone at the beginning of the simulation, allowing the DZS algorithm to derefine chunks of that volume right away on the first timestep.

\section{Summary and conclusions} \label{sec:conclusions}
In this paper, we presented implementations of the DZS algorithm in the codes for N-body simulations Arepo and Gadget4 across different cosmological models. This method, originally introduced in the Gadget3 code \citep{garaldi_dynamic_2020}, substantially improves the performance of large-scale cosmological simulations that produce a light-cone-like output, with a minimal impact on the accuracy of the results. This performance improvement is achieved by focusing computational resources inside the light cone by reducing the simulation resolution outside of it. The algorithm is also fully compatible with the gravitational treePM solver widely adopted in state-of-the-art codes, and only requires minimal modifications to their workflow. In fact, this high degree of compatibility allows the algorithm to work even in nonstandard cosmologies, like modified gravity (modeled in the code MG-Arepo through a computationally expensive multi-grid solver) and dynamic dark energy models (in the code PANDA-Gadget4), making it a very powerful performance-enhancing tool for constraining such models.

Enabling and validating the DZS approach in these beyond-$\Lambda$CDM codes yielded the following results:
\begin{itemize}
    \item Simulations performed with the DZS algorithm complete in significantly less time relative to their standard counterparts; in our testing setups, a large cubic volume  of $8192$ cMpc$/h$ on a side yields a run time saving of up to $\sim 50\%$. We also tested a smaller volume of $2048$ cMpc$/h$ on a side, which still enabled a saving of up to $\sim 25\%$ (Table~\ref{tab:times}). The volume size correlates with the performance gain, since the light cone crosses larger volumes sooner.
    \item Other than the volume size, employing different cosmological models also affects the performance. While our $\Lambda$CDM results generally agree with those of the original Gadget3 implementation, more computationally expensive cosmological models yield higher DZS-related performance gains (Table~\ref{tab:times}). For instance, modified $f(R)$ gravity setups lead to up to $\sim 10\%$ more computing time savings than $\Lambda$CDM or DS simulations.
    \item By comparing light-cone-like outputs from the same simulations performed with and without the DZS algorithm, we assessed its accuracy in reproducing such output. Relative differences in the light cone halo mass function (Figs.~\ref{fig:LCHMFlcdm} and \ref{fig:LCHMFother}), sky-projected mass maps (Fig.~\ref{fig:massmaps}), and angular power spectra of matter (Figs.~\ref{fig:pslcdm}, \ref{fig:psmediumHR}, \ref{fig:psmedium}, and \ref{fig:pslargeHR}) and weak lensing convergence (Fig.~\ref{fig:pskappamediumHR}) are typically of the order of $0.1\%$ or lower. Boosts and de-boosts in the matter power spectra due to nonstandard effects are also accurately reproduced by the DZS algorithm (Fig.~\ref{fig:psboost}). Such comparisons are in good agreement with observational forecasts for Stage-IV surveys, which further demonstrate that DZSs can help with the modeling and interpretation of their data.
    \item Our tests suggest that higher resolution leads to larger time savings, due to the more accurate, and therefore more computationally expensive, depiction of nonlinear matter clustering effects (Table~\ref{tab:times}). 
    To estimate the performance gain in large-scale simulations with state-of-the-art resolutions,
    we simulated a small volume at a much higher resolution than our default test setups, and approximated the effects of the DZS algorithm in large boxes through a rescaling of the light cone. This approach predicted run time savings of $\sim 65\%$ with a $8192$ cMpc$/h$ box (Fig.~\ref{fig:hires}), in agreement with the trend of our validation setups. This result improves up to an impressive $\sim75\%$ wall-clock time reduction for a larger $10240$ cMpc$/h$ box.
\end{itemize}

In conclusion, our implementation and validation work suggests that DZSs are a very valuable tool to address the technical challenges of the next generation of cosmological simulations. In particular, this technique enables a computationally cost-effective way to thoroughly constrain the unprecedented amount of data from current and upcoming surveys, across multiple cosmological models. 

\begin{acknowledgements}
We are very grateful to everyone that made this work possible. We thank Volker Springel for granting access to the developer version of Arepo, as well as providing extensive explanations and support on specific parts of the code. We also thank Baojiu Li, Christian Arnold, and César Hernandez Aguayo for allowing us to use MG-Arepo in our analysis. 
We acknowledge the Open Physics Hub project (hosted by the University of Bologna) for granting us access to their computational resources. We also gratefully acknowledge the Gauss Centre for Supercomputing e.V. (\url{www.gauss-centre.eu}) for funding this project by providing computing time on the GCS Supercomputer SuperMUC-NG at Leibniz Supercomputing Centre (\url{www.lrz.de}). In addition, we acknowledge the EuroHPC Joint Undertaking for awarding this project access to the EuroHPC supercomputer LUMI, hosted by CSC (Finland) and the LUMI consortium, through EuroHPC Development and Benchmark Access calls. We are grateful to Fulvio Ferlito for providing active and extensive support for the \texttt{dorian} package. RZ is funded by the European Union - NextGenerationEU, National Recovery and Resilience Plan (NRRP), Mission 4, Component 2, CUP J33C23002460002. EG is supported by the Kakenhi ILR 23K20035 grant. Finally, we thank the developers and maintainers of other Python packages used in this work, namely \texttt{numpy} \citep{walt_numpy_2011}, \texttt{scipy} \citep{virtanen_scipy_2020}, \texttt{h5py} \citep{collette_python_hdf5_2014}, \texttt{matplotlib} \citep{hunter_matplotlib_2007}, \texttt{healpy} \citep{zonca_healpy_2019} and \texttt{ligo.skymap} \citep{singer_bayesian_2016, singer_ligo_2016}.
\end{acknowledgements}

\bibliographystyle{aa}
\bibliography{references.bib}

\begin{appendix}
\onecolumn
\section{Angular matter power spectra of additional nonstandard simulations} \label{sec:A1}
\begin{figure*}[h!]
   \includegraphics[width=0.495\hsize]{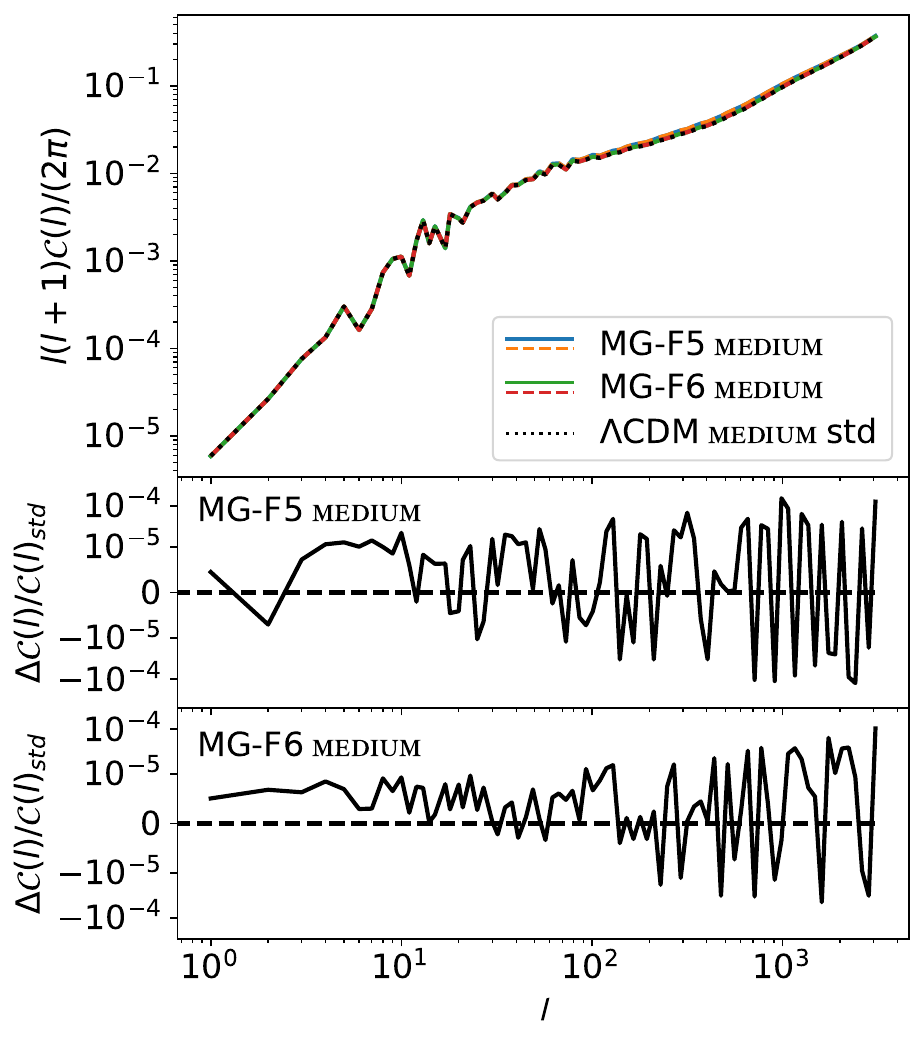}
   \includegraphics[width=0.495\hsize]{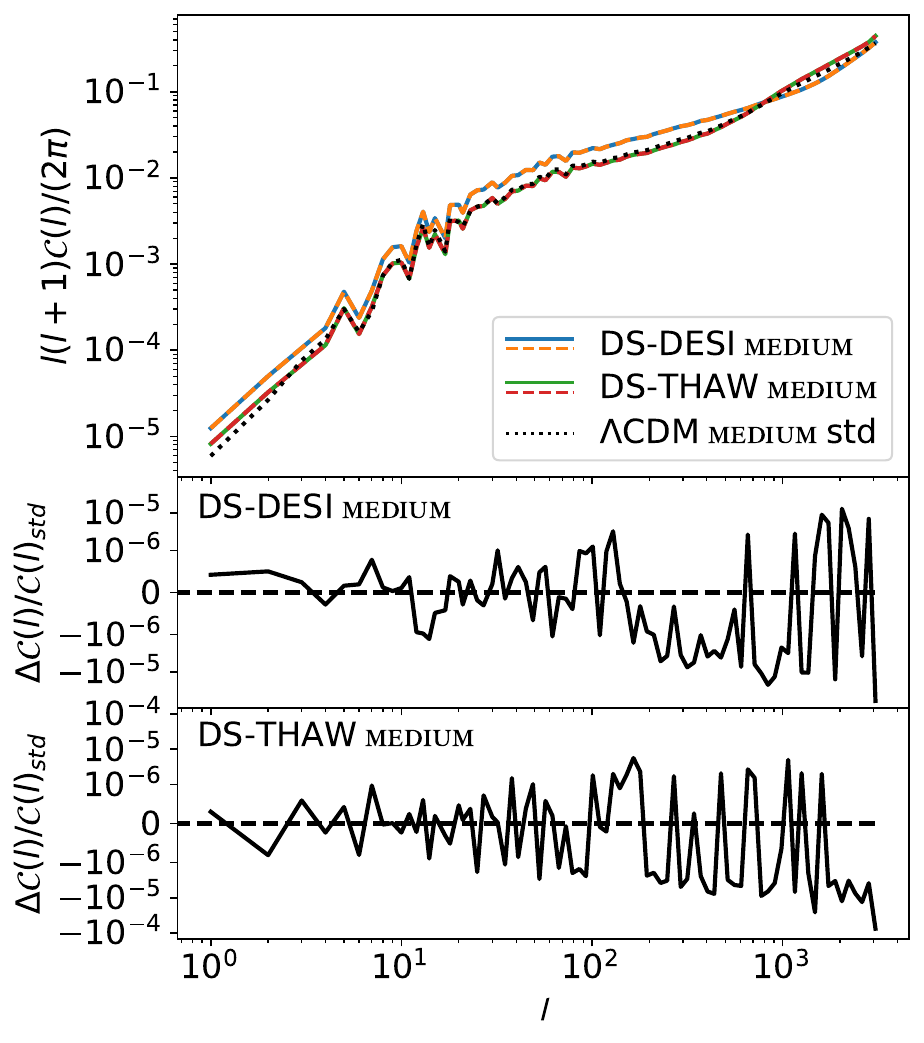}
      \caption{\textit{Left column}: Angular power spectrum ($\mathcal{C}(l)$) of the MG-Arepo \textsc{medium} simulations, for the MG-F5 and MG-F6 models (\textit{top}). Solid lines refer to \texttt{std} runs, dashed lines to \texttt{dzs} ones. The $\Lambda$CDM \texttt{std} case (dotted line) is overlaid as a reference. Relative difference plots are included for both MG models (\textit{middle} and \textit{bottom}, as labeled), with a dashed line indicating  $\Delta \mathcal{C}(l)=0$ for reference. \textit{Right column}: Same but for the DS-DESI and DS-THAW \textsc{medium} simulations.} 
         \label{fig:psmedium}
\end{figure*}
\begin{figure*}[h!]
   \includegraphics[width=0.495\hsize]{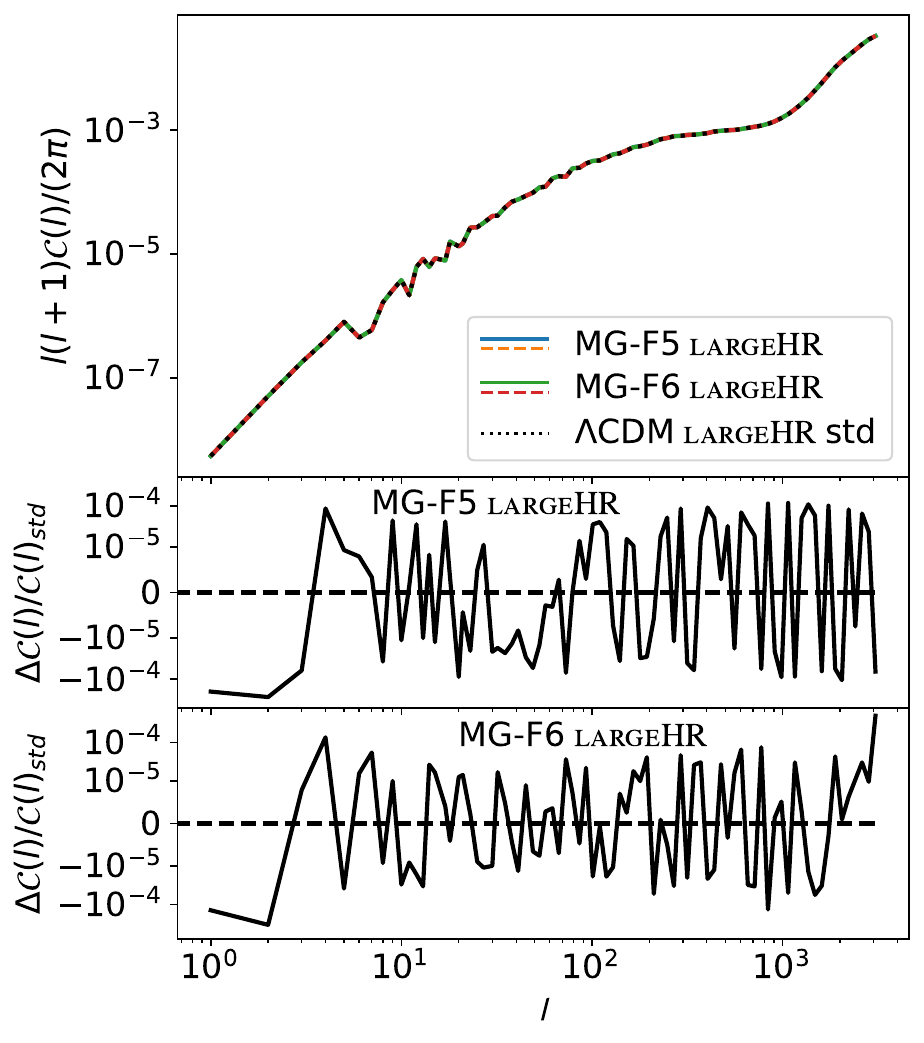}
   \includegraphics[width=0.495\hsize]{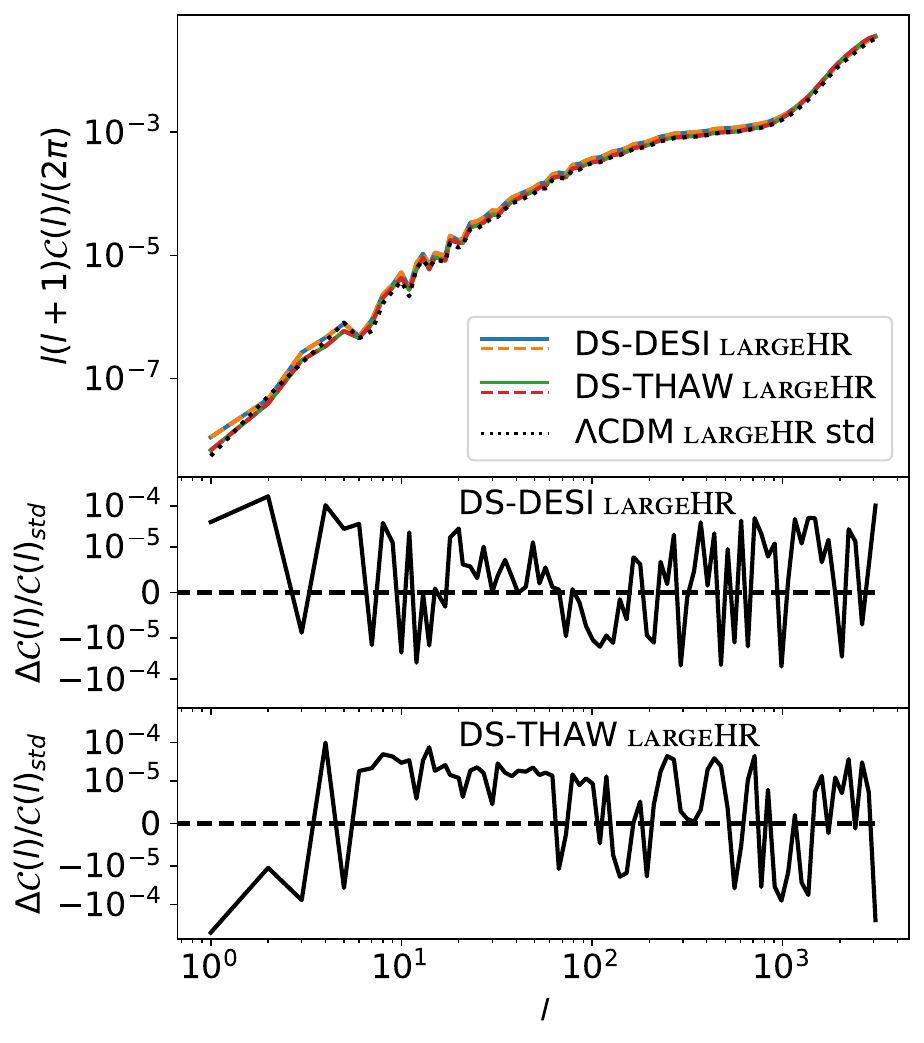}
      \caption{Same as Fig.~\ref{fig:psmedium} but for the \textsc{largeHR} simulations.} 
         \label{fig:pslargeHR}
\end{figure*}
\twocolumn
\noindent In this appendix we present angular matter power spectra from the beyond-$\Lambda$CDM \textsc{medium} and \textsc{largeHR} simulations, in Figs.~\ref{fig:psmedium} and \ref{fig:pslargeHR}, respectively. As in the \textsc{mediumHR} and $\Lambda$CDM simulations described in Sect. \ref{sec:hpix}, relative differences between \texttt{dzs} and \texttt{std} simulations remain very small, never exceeding $\sim 0.01\%$. Nonstandard features are still recognizable with respect to the reference standard model (dotted black line), especially in the \textsc{medium} setups (thanks to the narrower redshift range where output is produced). Even though a comparison between Figs.~\ref{fig:psmediumHR} and \ref{fig:psmedium} reveals that the accuracy does not always increase with increasing resolution (as already noted in Sect. \ref{sec:hpix}) these additional results remark the very high accuracy yielded by the DZS algorithm, across the different setups and models of our validation suite.

\section{Workload balance} \label{sec:wl}
While the DZS-induced performance gain observed in Sect. \ref{sec:performance} is significant, we assessed if there is room for further improvement, by checking how the workload balance\footnote{The workload balance provides an estimate of how well the work is distributed between parallel computing tasks. For each timestep, it is defined as the maximum time, among all tasks, taken to evaluate tree gravitational interactions, divided by the average value.} changes when the DZS algorithm is operating. Indeed, we expect that the presence of multiple resolutions in a simulation, as well as the overall smaller number of particles, will make it more difficult to evenly distribute the work and particle load among parallel tasks. 

\begin{figure}[t]
   \resizebox{\hsize}{!}{\includegraphics{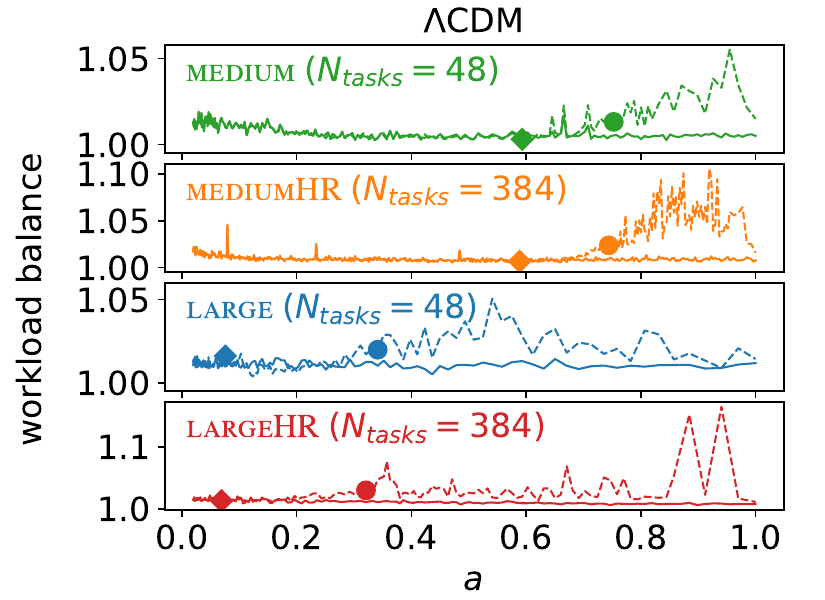}}
      \caption{Workload balance for all the $\Lambda$CDM simulations performed in this work as a function of the scale factor ($a$). The number of parallel tasks $N_{tasks}$ is included for each setup. Solid lines refer to \texttt{std} runs, dashed lines to \texttt{dzs} ones. The diamond symbol marks the time at which the light cone enters the simulation volume and the DZS algorithm starts merging particles. The circular symbol shows when the number of particles becomes lower than half of the initial value as a result of the merging operation.} 
         \label{fig:wl}
\end{figure}

Figure~\ref{fig:wl} shows the workload balance as a function of the scale factor $a$ for all  $\Lambda$CDM twin simulations, with solid lines referring to \texttt{std} runs and dashed lines to \texttt{dzs} ones. From the moment in which the light cone enters the simulation volume (indicated by the diamond marker in Fig.~\ref{fig:wl}) and the DZS algorithm starts to merge particles, the balance in \texttt{dzs} runs diverges from the ideal value of $1$, going as high as $\sim 1.1$ in the \textsc{mediumHR} and \textsc{largeHR} simulations. The effect is especially prominent after the number of particles in DZSs drops below half of the initial value (a time marked by the circle in the figure). While there is still room for improvement, for example through appropriate weighting in the domain decomposition or a change in the number of computing tasks mid-simulation \citep[the so-called ``DZS special stop'' in][]{garaldi_dynamic_2020}, this result represents only a slight imbalance. In fact, part of the observed imbalance could also be caused by machine fluctuations, as in the case of the \textsc{large} simulations, where the balance lies noticeably above the ideal value also in the \texttt{std} run.

\begin{figure*}[ht!]
   \includegraphics[width=0.495\hsize]{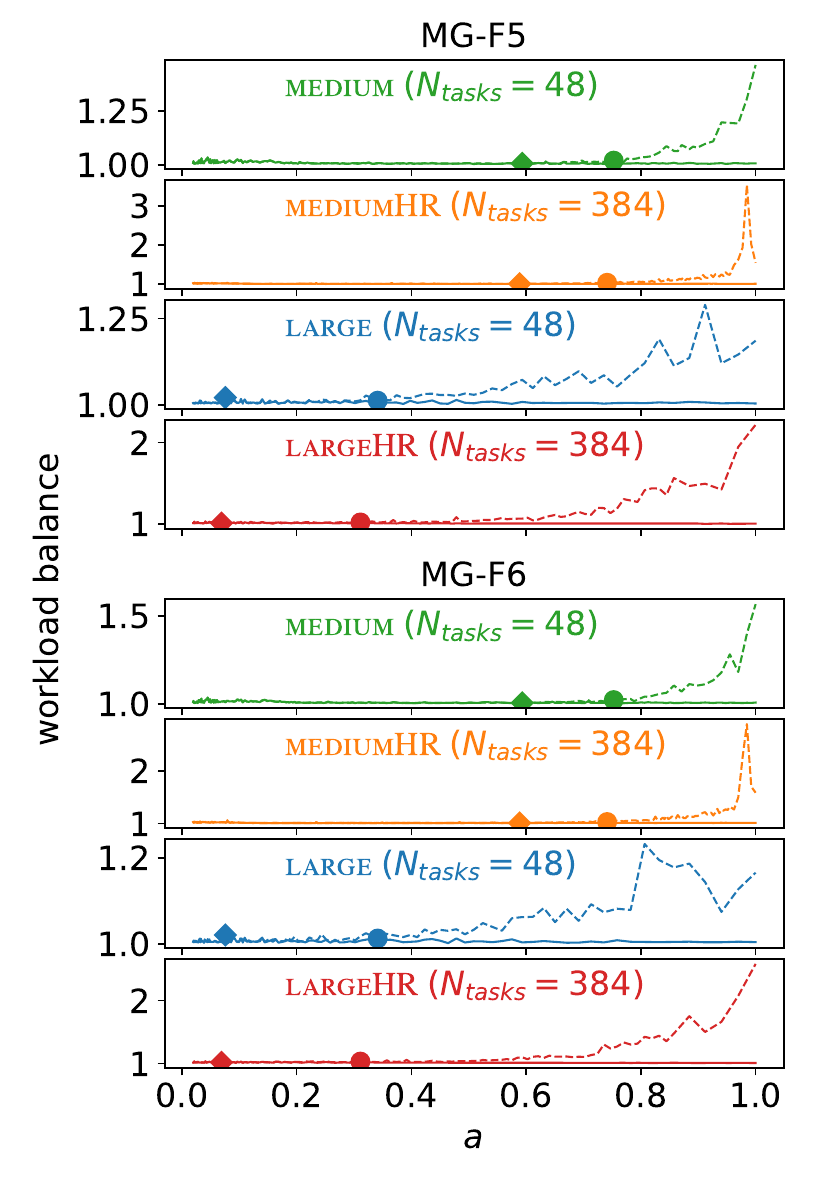}
   \includegraphics[width=0.495\hsize]{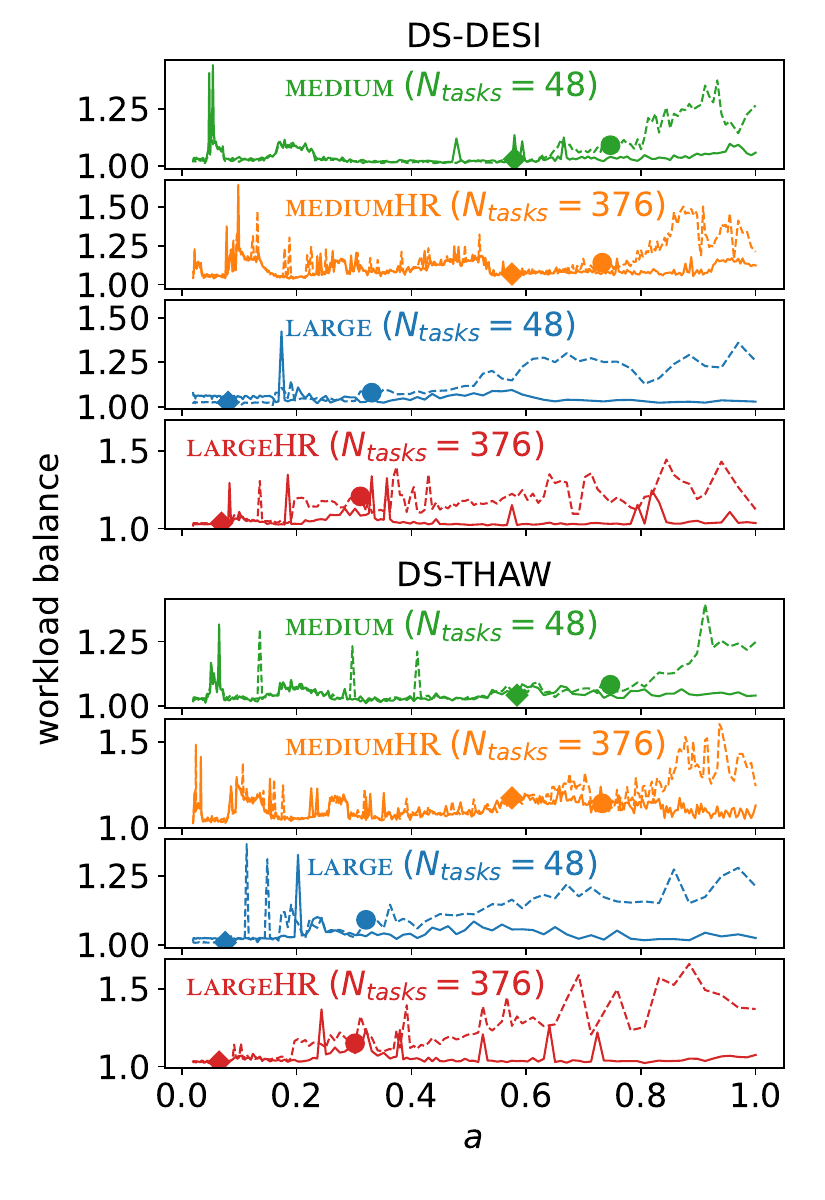}
      \caption{\textit{Left column}: Workload balance of all of the MG-F5 (top) and MG-F6 (bottom) simulations tested as a function of the scale factor ($a$). Solid lines refer to \texttt{std} runs, dashed lines to \texttt{dzs} ones. The number of parallel tasks $N_{tasks}$ is included for each setup. The diamond symbol marks the time at which the light cone enters the simulation volume and the DZS algorithm starts merging particles. The circular symbol shows when the number of particles becomes lower than half of the initial value as a result of the merging operation. \textit{Right column}: Same but for the DS-DESI and DS-THAW simulations.} 
         \label{fig:wlother}
\end{figure*}

The workload balance for beyond-$\Lambda$CDM models is shown in Fig.~\ref{fig:wlother}. The lower resolution MG setups, that is, the \textsc{medium} and \textsc{large} runs in the left column of the figure, behave relatively well, with peaks up to $\sim 1.5$ in the \texttt{dzs} MG-F6 \textsc{medium} simulation. Their higher-resolution counterparts, however, peak at values of $2$ or even $3$, as in the case of the MG-F5 \textsc{mediumHR} setup. While such values are limited to a handful of timesteps, they will need to be thoroughly investigated to improve the performance of DZSs in modified gravity setups. We leave such an investigation to a potential follow-up analysis, as the observed run time saving is already satisfactory ($\gtrsim 50\%$ in the largest volume). Dark scattering simulations (right column of Fig.~\ref{fig:wlother}) align with the lower-resolution MG setups, with \texttt{dzs} runs going as high as $\sim 1.5$ in workload balance. In fact, the most prominent feature arising in DS runs (both \texttt{std} and \texttt{dzs}) is the very narrow spikes resulting from the use, in Gadget4, of the hybrid shared-distributed memory variant of MPI mentioned in \ref{sec:gadget4}. Normally, each parallel MPI task is assigned its own memory and needs to communicate with other tasks to exchange contents of that memory. Conversely, Gadget4 uses a version of MPI where memory is shared within tasks on the same node (mirroring the real, physical structure of individual computing nodes), and explicit communications are performed only between two or more nodes. This reduces memory overhead, but also introduces potential race conditions and inconsistent ordering of operations performed by different tasks. These break binary invariance of multiple reruns, which would hinder the accuracy tests performed in Sect. \ref{sec:validation}. Gadget4 allows us to retain the invariance by forcing tasks to wait for others to access shared memory first, introducing sudden and temporary imbalances in the workload. Nonetheless, the DZS-induced gradual imbalance is still clearly recognizable. In conclusion, in both the MG and DS models, the DZS algorithm yields significant workload imbalances, which call for dedicated optimization techniques to be kept in check. We note that lowering the workload imbalance could potentially lead to even larger performance gains than those reported in Table~\ref{tab:times}.

\end{appendix}

\end{document}